\documentclass[12pt]{iopart}

\usepackage{iopams}
\expandafter\let\csname equation*\endcsname\relax

\expandafter\let\csname endequation*\endcsname\relax
\usepackage{physics}
\usepackage[x11names]{xcolor}
\usepackage{graphicx}
\usepackage{color}
\usepackage{lipsum}
\usepackage{hyperref}
\usepackage{cite}
\usepackage[normalem]{ulem}
\usepackage{comment}
\usepackage{xcolor}
\usepackage{mathrsfs}
\usepackage{dsfont} 
\usepackage{bm}
\usepackage{ulem}

\hypersetup{
    colorlinks=true,
    linkcolor=blue,
    citecolor=blue,
    urlcolor=blue
}

\renewcommand{\vec}{\mathbf}

\begin{document}
 
\title[]{Time irreversibility and entropy production in non-Hermitian Model A field theories}

\author{Matthias Carosi$^{1,2}$\footnote{These authors have contributed equally to this work.}, Ot Garcés$^{3,4}$\footnotemark[1], Adrià Garcés$^{3,4}$, Demian Levis$^{3,4}$}

\address{$^1$Physik Department T70, Technische Universit\"at M\"unchen,\\James-Franck-Str., D-85748 Garching, Germany}
\address{$^2$International Centre for Theoretical Physics Asia-Pacific (ICTP-AP), \\University of Chinese Academy of Sciences, 100190 Beijing, China}
\address{$^3$Computing and Understanding Collective Action (CUCA) Lab, Departament de Física de la Matèria Condensada, Universitat de Barcelona, Mart\'i i Franqu\`es 1, E08028 Barcelona, Spain}
\address{$^4$University of Barcelona Institute of Complex Systems (UBICS),  Martí i Franquès 1, E08028 Barcelona, Spain}

\ead{matthias.carosi@tum.de}
\vspace{10pt}
\begin{indented}
\item[]\today
\end{indented}
\begin{abstract}
 
We develop a systematic framework to quantify irreversibility in scalar Model A field theories with a generic non-Hermitian term driving the dynamics. Using the stochastic path-integral formalism, we perform a controlled small-noise expansion, allowing the computation of the entropy production rate (EPR) and violations of the fluctuation-dissipation theorem (FDT). We show that the local EPR  is entirely determined by the anti-Hermitian part of the linearised Langevin equation. Around steady states, the non-Hermitian component produces linear corrections to FDT violations and contributes quadratically to the EPR. As an illustration of the applicability of our approach, we analyse a minimal non-Hermitian extension of the Ginzburg–Landau $\psi^4$ theory describing a non-reciprocal Ising model at coarse-grained scales, for which we obtain explicit expressions of the local EPR, showing that it localises at interfaces in non-uniform states. Our results provide a general characterisation of TRS breaking in non-Hermitian scalar field theories.
\end{abstract}

%
\vspace{2pc}

\vspace{1em}
 
\maketitle
 
\tableofcontents

\section{Introduction}
\label{sec:intro}

Time-reversal symmetry (TRS) is the defining feature of equilibrium. The stochastic dynamics of a system coupled to an equilibrium bath satisfies detailed balance: trajectory probabilities are invariant under time reversal,  ultimately constraining both steady-state distributions and dynamical correlations. However, large-scale phenomena as we observe them typically occur under non-equilibrium conditions, following spontaneous evolutions that clearly look time irreversible \cite{lebowitz1993boltzmann}.   

In a large variety of non-equilibrium systems, such as in active matter, 
energy injection occurs locally at the level of the microscopic degrees of freedom \cite{byrne2022time}.
While at the microscopic particle level, TRS breaking is explicit in such systems, identifying its signatures at large collective scales remains far from obvious. Indeed, irreversible microscopic dynamics may lead to emergent macroscopic behaviour that resembles equilibrium phenomenology; a system endowed with a microscopic dynamics breaking detailed balance might not exhibit any clear trace of "non-equilibriumness" at coarse-grained scales \cite{Elsen18, Li20}. 
A salient example of such difficulties to identify large-scale signatures of TRS breaking in active matter is Motility-Induced Phase Separation (or MIPS) \cite{Tailleur08, Cates15}. Such a phenomenon emerges from the competition between excluded volume interactions and persistent motion (or persistent noise), triggering a phase separation, very much like a system of attractive passive particles cooled below their liquid-gas condensation temperature \cite{Solon18}.  
One has to deeply look at the structure and kinetics of these condensates to eventually distinguish active from passive phase separation   \cite{digregorio2018full,  caporusso2020motility, shi2020self, caporusso2023dynamics, cates2025active}. 
This illustrates a broader fundamental question: how can one systematically quantify irreversibility at macroscopic scales?

Two complementary approaches, or metrics, have been developed in the literature. 
One relies on an (informatic) entropy production rate (EPR) defined at the trajectory level \cite{Lebowitz99, maes2003time, gaspard2004time, Seifert12,fodor2016far, fodor2022irreversibility}. Despite many recent efforts, the systematic computation of such EPR in interacting many-body systems remains technically challenging, both from a computational, analytical, and experimental viewpoint \cite{cocconi2020entropy,petrelli2020effective,cengio2021fluctuation, ro2022model, byrne2022time, semeraro2024entropy, hecht2024define, di2024variance, Pruessner25}.
A second approach 
resorts to the violations of the Fluctuation-Dissipation theorem (FDT) \cite{cugliandolo2011effective}. 
This approach has indeed been proven very fruitful in the context of slowly relaxing systems \cite{cugliandolo1997energy, berthier2002nonequilibrium}, and it is experimentally accessible  \cite{grigera1999observation}. 
These two metrics, FDT violations and EPR, can be bridged together via Harada–Sasa relations \cite{Harada05},  providing a unified characterisation of TRS breaking.  {More recently, complementary approaches based on vorticity and cycle affinity have also been introduced to quantify irreversibility in stochastic dynamics \cite{JO23,JO24,JO25}.
}

A useful (simplified) approach to quantify irreversibility in many-body systems on general grounds, is through the analysis of  coarse-grained field descriptions \cite{hohenberg1977theory,tauber2014critical}.
Although a variety of non-equilibrium dynamical field theories have been considered so far,  we shall focus our discussion here on  {real} scalar fields \cite{cates2019active, CN17, Markovich21, LC21, Loos23, SuchanekPRE2023, SuchanekPRL2023,SuchanekPRE2024,paoluzzi2024noise,johnsrud2025,Golestanian25}.  
In this context, Nardini et al. \cite{CN17} extended the definition of EPR to field theories, and computed it numerically from stochastic trajectories of a field theory, aiming at first to describe MIPS  (the so-called Active Model B). They derived a Harada--Sasa type of relation and showed that the EPR can be decomposed locally, concentrating on the interfaces generated through MIPS. 
Subsequently, Li and Cates \cite{LC21} provided a systematic methodology  to compute EPR in scalar Langevin field theories via a small-noise expansion, under the assumption of Hermiticity of the linear fluctuation operator around steady states.

More recently, non-Hermitian dynamical field theories have emerged in the context of non-reciprocal interactions \cite{fruchart2026nonreciprocal}. In particular, Suchanek, Kroy, and Loos extended the approach of Li and Cates, and studied EPR in non-reciprocal Cahn–Hilliard dynamics (Model B with asymmetric couplings \cite{you2020nonreciprocity, saha2020scalar}), highlighting the role of parity–time symmetry breaking and irreversible mesoscale fluctuations \cite{SuchanekPRE2023, SuchanekPRL2023,SuchanekPRE2024}.
They derived general results for scalar fields with conserved dynamics and observed how the EPR is generated by the non-Hermitian part of the linearised Langevin evolution\,\cite{SuchanekPRE2024}.
Parallel developments by Johnsrud and Golestanian \cite{johnsrud2025, Golestanian25} have considered a broad class of models (including both conserved and non-conserved order-parameter dynamics) for which the EPR and the violations of the FDT can be computed via perturbation expansions in the couplings, using the stochastic path-integral formalism.
Non-Hermitian field theories have also attracted considerable attention across several areas of physics, even beyond the study of statistical systems. 
They arise naturally in effective descriptions of dissipative or open quantum systems, where non-Hermitian generators appear in Lindblad dynamics\,\cite{PRXQuantum.4.030328,gwzr-cjp6}, and have also been explored in relativistic quantum field theory and cosmology as extensions of conventional Hermitian models\,\cite{PhysRevD.102.125030,Millington:20228a}.

In this work, we develop a systematic framework to quantify the impact of a generic non-Hermitian term in  {the linearised dynamics of a} scalar field  with non-conserved dynamics (Model A) on time-reversal symmetry breaking.
Using the path-integral formalism, we perform a controlled small-noise expansion at the level of the generating functional, which provides a complete statistical description of fluctuations around deterministic  states.
Within this approach, FDT violations can be computed exactly in the linearised theory and related to the EPR through generalised Harada–Sasa relations.

Extending the method of Li and Cates \cite{LC21} to non-Hermitian operators, we show that the local EPR can be expressed solely in terms of the anti-Hermitian component of the linearised Langevin equation. 
In particular, we demonstrate that, around steady deterministic solutions, the anti-Hermitian part generates linear corrections to FDT violations while contributing quadratically to the EPR. 
This includes situations where the underlying dynamics is Hermitian, but the expansion around a steady state leads to a non-Hermitian linearised evolution.
Our results for the entropy production in uniform backgrounds are consistent with those obtained in Ref.\,\cite{SuchanekPRE2024}. Here, we go beyond this setting by establishing a direct relation between EPR and FDT violations through a generalised  {local} Harada--Sasa relation, and by extending the analysis to spatially non-uniform backgrounds.
To illustrate these results, we consider a minimal non-Hermitian extension of the Ginzburg–Landau $\psi^4$ theory, motivated by coarse-grained descriptions of single-species non-reciprocal spin systems \cite{Garc_s_2025, akritidis2026fate}. For this model, we obtain analytical expressions for the local EPR in both uniform and non-uniform states. The calculations show that entropy production vanishes in symmetric phases and becomes localised at interfaces in heterogeneous states, consistent with previous numerical observations in active field theories \cite{CN17, paoluzzi2024noise}.
 {Interestingly, the local EPR vanishes exactly at the centre of the interface, where the symmetry is restored.}

The article is organised as follows. 
In section \ref{sec:functional} we introduce a generic scalar field theory with Model A dynamics and non-Hermitian forces. In section \ref{sec:FDT} we  {briefly review the response-field formalism}, a path integral  {framework}  to access correlation and response functions.
We then perform a small-noise expansion around a stationary uniform solution of the field equations, and show that the violations of the FDT arise as soon as non-Hermitian forces are present.
FDT violations are quantified by a fluctuation-dissipation difference that can be related to the EPR via a  {local} Harada--Sasa relation. 
Then, in section \ref{sec:langevin}, we derive a general expression of the local EPR directly from the Langevin equation of motion of the model. 
We apply the formalism to the case of a self-advecting non-Hermitian force, as arising in spin models with vision cone interactions, among others, and derive an explicit expression of the local EPR in both a uniform and non-uniform phase. The latter calculation shows that the local EPR is localised at the boundaries between ordered regions. 
We finally close the article in section \ref{sec:conclusion} with a summary of our findings, a general discussion of its implications, and some perspectives for future work.

\section{Non-Hermitian Model A dynamics}
\label{sec:functional}

We consider a system described by a single real scalar field~$\psi$ evolving under Model A (non-conserved order parameter) dynamics in $d$ spatial dimensions \cite{hohenberg1977theory}.
The field obeys the overdamped Langevin equation
\begin{equation}
    \label{eqn:Langevin_eq}
    \dot\psi (\textbf{x},t)=  \Gamma F[\psi](\textbf{x},t) + \sqrt{2\gamma}\xi(\textbf{x},t) \,,
\end{equation}
where the dot stands for time derivative, $\Gamma$ is a constant relaxation rate or mobility, $F[\psi]$ is a deterministic force, $\gamma$ characterises the noise strength and $\xi$ is Gaussian {unit white} noise with 0 mean and unit variance, namely 
\begin{align}
    {\langle \xi(\textbf{x},t)\rangle = 0, \quad
    \langle \xi(\vec x,t) \xi(\vec x', t') \rangle =  \delta(\textbf{x}-\textbf{x}')\delta(t-t')}.
\end{align}
We also assume that the relaxation rate $\Gamma$ and the noise amplitude $\gamma$  are related by the usual Einstein relation $\Gamma =\beta \gamma$.

 {The goal of the present work is to quantify departures from equilibrium in  real scalar field theories that are driven by a non-Hermitian force when linearised about stationary backgrounds. 
The non-Hermiticity of the linearised dynamics is identified by the fluctuation operator about the given stationary state, failing to be self-adjoint with respect to the $L^2$-scalar product. 
 {In section \ref{sec:FDT_viol_uniform_background} we show that} when  {linearising} about a uniform stationary state that preserves translational  invariance, non-Hermiticity is equivalent to the breaking of parity symmetry.
However, such equivalence is not generic, and no longer holds for non-uniform backgrounds. We provide a clear example in section~\ref{sec:domain_wall}, where we look at fluctuations around a domain wall stationary state and show that the non-equilibrium character of the theory is manifestly induced by a non-Hermitian and yet parity-even driving force.
It is non-Hermiticity that tracks the deviation from  equilibrium.
}

 {A non-Hermitian dynamics, for example, may arise when}
coarse-graining a microscopic theory that includes non-reciprocal interactions, such as the non-reciprocal Ising model studied in Ref \cite{Garc_s_2025}.
In said model, the Ising spins have a vision-cone dependent interaction, that in the continuum leads to a self-advection term contributing to the deterministic force with terms of the form $\Gamma F[\psi] \supset \: \lambda \psi (\mathbf{v} \cdot \nabla) \psi$.
Here, $\mathbf{v}$ is the vision cone vector that is fixed by the microscopic model,  {which causes the field to self-advect in a fixed direction of space}.
 {After linearising the theory, this yields a term with a single derivative operator, which is explicitly anti-Hermitian.}

Non-Hermiticity can arise in multi-species models too, such as the non-reciprocal Cahn-Hilliard model \cite{you2020nonreciprocity, saha2020scalar,SuchanekPRE2023, Golestanian25}.
 {
Also the Kardar-Parisi-Zhang (KPZ) equation \cite{kardar1986dynamic} offers a well-studied example, where the linearised theory expanded around a phase boundary configuration renders the fluctuation operator non-Hermitian.}

With no loss of generality, the force can be split into conservative and non-conservative components, namely
\begin{align}
    F[\psi](\vec x, t) = - \frac{\delta H[\psi]}{\delta \psi(\vec x, t)}  + \mathscr{F}[\psi](\vec x, t).
\end{align} 
The conservative force term is given by the functional derivative of a Ginzburg--Landau -type functional $H[\psi]$.  
The non-conservative force term $\mathscr{F}[\psi]$ gathers all the remaining terms that cannot be written as a functional derivative of a free energy functional. 
It is precisely the fact that these driving forces cannot be written as functional derivatives of any 'free-energy' function that gives rise to steady-states breaking TRS, as already discussed in  Refs.\,\cite{CN17, SuchanekPRL2023,SuchanekPRE2023, Golestanian25}.  {Note, however, that the decomposition into conservative and non-conservative components is not unique \cite{CN17}, though this does not affect any of the results below}.

\section{Quantifying irreversibility via FDT violations}
\label{sec:FDT}

Our first approach to quantify TRS breaking is through the violations of the FDT. To do so, we construct a  generating functional for the correlation and response functions  in terms of path integrals using the  Martin-Siggia-Rose-Janssen-De Dominicis (MSRJD), or response-field, formalism \cite{MSR_73, de2022path}, 
which is particularly well-suited for the computation of response functions.
We then perform a small-noise \cite{gardiner2010stochastic, van2011stochastic} or semi-classical expansion about uniform stationary field configurations, corresponding to the zero-noise steady solutions of Eq.\,\eqref{eqn:Langevin_eq}. As we shall see, this allows us to identify analytically how 
 {parity-breaking} forces break the equilibrium FDT.

 {While readers experienced with the response-field formalism may find some of the intermediate calculations rather standard, we include them for pedagogical reasons. Their explicit presentation highlights that all subsequent results follow naturally from a single semi-classical field-theoretic framework, therefore providing a unified treatment of the problem. Readers already familiar with these techniques may wish to proceed directly to Sec. \ref{sec:small-noise-exp}.}

\subsection{Response-field formalism and generating functional}

Using standard field-theoretical methods \cite{tauber2014critical}, the generating functional associated with the Langevin equation~\eqref{eqn:Langevin_eq} within the MSRJD formalism is given by 
\begin{equation}
    \label{eqn:generating_functional}
    Z[J,\widetilde{J}] = \: \mathcal{N} \int \mathcal{D}[\psi,\widetilde{\psi}]\: \exp \left[-\mathscr{A}[\psi,\widetilde \psi] + \int_{\textbf{x},t} \left(\widetilde{J} (\textbf{x},t)\psi(\textbf{x},t) + J (\textbf{x},t)i\widetilde\psi(\textbf{x},t)\right)\right] \,,
\end{equation}
where the shorthand notation   $\int_{\textbf{x},t}=\int d^d\textbf{x}\,dt$ for the integrals over all space and time has been used. The two external sources   $J$ and $\widetilde J$ are coupled to the field $\psi$ and response field $\widetilde\psi$ respectively.  The normalisation factor  $\mathcal{N}$ is chosen such that $Z[0,0]=1$. The dynamic action reads  
\begin{equation}
    \label{eqn:MSR_action}
     {\mathscr{A}[\psi,\widetilde\psi] = \: \int_{\textbf{x},t}  i\widetilde\psi(\textbf{x},t)   \left(\dot\psi(\textbf{x},t) - \Gamma F[\psi](\textbf{x},t) - \gamma i \widetilde{\psi}(\textbf{x},t)\right)}
\end{equation}

The expectation value of any observable, $\mathcal{O}[\psi, \widetilde{\psi}]$, is obtained by averaging with respect to the statistical weight $e^{-\mathscr{A}[\psi, \widetilde{\psi}]}$, namely
\begin{equation}
    \langle \mathcal{O} [\psi,\widetilde\psi] \rangle = \: \mathcal{N} \int \mathcal{D}[\psi,\widetilde\psi] \, \mathcal{O} [\psi,\widetilde\psi] \, e^{- \mathscr{A}[\psi,\widetilde\psi]} \,.
\end{equation}
Therefore, the two-point correlation and response functions defined by $C(\textbf{x},\textbf{x}';t,t') = \langle \psi(\textbf{x},t)\psi(\textbf{x}',t')\rangle$ and $G(\textbf{x},\textbf{x}';t,t') = \langle \psi(\textbf{x},t)i\widetilde{\psi}(\textbf{x}',t')\rangle$ are directly computed from the generating functional as
\begin{subequations}
\begin{align}
      C(\mathbf x, \mathbf x'; t,t') = \: & \frac{\delta^2 Z[J,\widetilde J]}{\delta \widetilde J(\mathbf x, t) \delta\widetilde J(\mathbf x', t')} \Bigg\rvert_{\substack{J=0\\\widetilde J=0}} \,, \label{eqn:correlation_from_Z}\\
      G(\mathbf x, \mathbf x'; t,t') = \: & \frac{\delta^2 Z[J,\widetilde J]}{\delta \widetilde J(\mathbf x, t) \delta \vphantom{\widetilde J} J(\mathbf x', t')} \Bigg\rvert_{\substack{J=0\\\widetilde J=0}} \,. \label{eqn:response_from_Z}
\end{align}
\end{subequations}
Integrating out the response field $\tilde{\psi}$ from Eq.\,\eqref{eqn:generating_functional} yields the Onsager-Machlup path integral \cite{onsager1953fluctuations}. In this work, we choose to use the MSRJD formalism, as it provides more direct access to response functions.

\subsection{Time-reversal symmetry as an equilibrium symmetry}

Equilibrium dynamics can be understood by the invariance of the dynamic action $\mathscr{A}$ under time-reversal \cite{Aron_2010}, defined by its action on the field and the response field, 
\begin{subequations}
\begin{align}
    \psi (\mathbf x, t) \xrightarrow{\mathcal{T}} \: & \psi(\mathbf x,-t) \,, \label{eqn:psi_under_T}\\
    i\widetilde\psi(\mathbf x, t) \xrightarrow{\mathcal{T}} \: & i\widetilde\psi(\mathbf x, -t) - \gamma^{-1}\dot\psi(\mathbf x, -t) \,. \label{eqn:psi_tilde_under_T}
\end{align}
\end{subequations}
Under  $\mathcal{T}$,  the action $\mathscr{A}[\psi, \widetilde{\psi}]$ thus transforms as
\begin{align}\label{eq:broken_invariance_action}
    \mathscr{A}[\mathcal{T}\psi, \mathcal{T}\widetilde{\psi}] = \mathscr{A}[\psi, \widetilde{\psi}] + \beta \int_{\textbf{x},t} \dot{\psi}(\textbf{x},t) \left ( -\frac{\delta H[\psi]}{\delta \psi (\textbf{x},t)} + \mathscr{F}[\psi](\textbf{x},t)\right). 
\end{align}
From Eq.~\eqref{eq:broken_invariance_action}, the change of the action under the time-reversal operation $\mathcal{T}$ amounts to two contributions: one associated with conservative forces $-\delta_\psi H[\psi]$ and one with the non-conservative driving $\mathscr{F}[\psi]$.
In the absence of non-conservative forces, i.e. when $\mathscr{F}[\psi]=0$, the change in the action amounts to a boundary term $\int_{\vec x,t} \dot\psi\, \delta_\psi H = \int_{t} d H / d t$.
In  steady conditions,
this contribution  {is compensated by the distribution of initial conditions when sampled from the correct Boltzmann distribution \cite{Aron_2010}},
and the action $\mathscr{A}$ is $\mathcal{T}$-invariant. 
For any observable $\mathcal{O}[\psi,\widetilde{\psi}]$, such a symmetry implies  the Ward-Takahashi identity
\begin{align}\label{eq:wt-equilibrium}
    \Big \langle \mathcal{O}[\mathcal{T}\psi, \mathcal{T}\widetilde{\psi}]\Big \rangle = \Big\langle \mathcal{O}[\psi, \widetilde{\psi}]\Big \rangle \,.
\end{align}
The Ward-Takahashi identity encodes the full set of equilibrium relations between correlations and responses\cite{Aron_2010}. 
For instance, the choice $\mathcal{O}[\psi, \widetilde{\psi}] = \psi(\vec x, t)i\widetilde{\psi}(\vec x', t')$ yields, after relabelling $t\leftrightarrow t'$ and exploiting time-translation invariance, the celebrated fluctuation-dissipation theorem
\begin{align}
    G(\textbf{x},\textbf{x}';t-t')-G(\textbf{x},\textbf{x}';t'-t)+\gamma^{-1}\partial_t C(\textbf{x},\textbf{x}';t-t') = 0.
\end{align}

\noindent
When non-conservative forces $\mathscr{F}[\psi]$ are present, the action is no longer $\mathcal{T}$-invariant, even in the long-time limit. 
Therefore, the broken invariance of the dynamic action in the steady regime prompts the definition of a two-point function $\Delta(\textbf{x},\textbf{x}'; t,t')$, the fluctuation-dissipation (FD) difference, which quantifies the violation of the FDT, namely
\begin{align}
    \label{eqn:def_FDT_violation}
    \Delta(\mathbf x, \mathbf x'; t,t') \equiv \: & G(\mathbf x, \mathbf x';t-t') - G(\mathbf x, \mathbf x';t'-t) 
    + \gamma^{-1}\partial_t C(\mathbf x,\mathbf x'; t-t') \,.
\end{align}
A hallmark of the departure from equilibrium is $\underline{\Delta}\neq 0$.

\subsection{Small-noise expansion about a stationary solution}\label{sec:small-noise-exp}

 {In the limit of small noise, violations of the FDT can be quantified explicitly.}
This is possible because the small-noise expansion effectively linearises the theory, allowing direct access to the correlation and response propagators at leading order   in the fluctuations around the zero-noise solutions, from which we derive an explicit expression for the FD difference $\underline{\Delta}$.  {}

In order to bring the dynamic action $\mathscr{A}[\psi, \widetilde{\psi}]$ into the canonical form, we re-scale the response field $i\widetilde{\psi} \mapsto i\widetilde{\psi}/\gamma$. 
This is convenient for a proper saddle-point computation of the generating functional in Eq.\,\eqref{eqn:generating_functional}.
We also re-scale the external source $J\mapsto \gamma J$, so that the noise amplitude $\gamma$ only appears as a pre-factor of the action.
Note that this requires updating the definition~\eqref{eqn:response_from_Z} of the response function $\underline{G}$ in terms of derivatives of the generating functional.
After the re-scaling, the generating functional $Z[J,\widetilde{J}]$ reads 
\begin{equation}
    \label{eqn:generating_functional_canonical}
    Z[J,\widetilde{J}] = \: \mathcal{N} \int \mathcal{D}[\psi,\widetilde{\psi}]\,\exp \left (-\frac{1}{\gamma} \mathscr{A}[\psi,\widetilde \psi] + \widetilde{J}^\dagger\psi + J^\dagger i\widetilde\psi\right) \,,
\end{equation}
where we absorb the Jacobian of the re-scaling into the irrelevant normalisation constant $\mathcal{N}$.
 {Here and in the following, we adopt a short-hand compact notation from functional analysis $v_1^\dagger v_2 = \:  \int_{\textbf{x},t}v_1^*(\mathbf x, t) v_2(\mathbf x, t)$ and  $v_1^\dagger \, \underline{A} \, v_2 = \:  \int_{\textbf{x},t} \int_{\textbf{x}',t'}  v_1^*(\mathbf x, t) \, A(\mathbf x, \mathbf x'; t,t')\, v_2(\mathbf x', t') $  with space-time integrals understood implicitly. Here, $*$ is the complex conjugation. 
The adjoint of a quadratic form is defined accordingly, via complex conjugation and the simultaneous exchange of space-time arguments $ Q^\dagger(\mathbf x, \mathbf x'; t,t') = \: Q^* (\mathbf x', \mathbf x; t', t)$.
While in coordinates $(\mathbf x, t)$ all the quantities we manipulate are manifestly real, this is not the case when working in some other basis, e.g. in Fourier space.
Conveniently, all expressions written in this compact notation are basis independent, so that the results carry on to any representation we wish to use.
}

In the small-noise limit, the functional integral~\eqref{eqn:generating_functional_canonical} for vanishing external sources $\widetilde J = J = 0$ is dominated by the saddle-point of the action, namely 
\begin{subequations}
\begin{align}
    \frac{\delta\mathscr{A}[\psi,\widetilde{\psi}]}{\delta i\widetilde\psi^\dagger} = \:  {\dot\psi - \Gamma F[\psi] - 2\,i\widetilde\psi} = \: & 0 \,, \\
    \frac{\delta\mathscr{A}[\psi,\widetilde\psi]}{\delta\psi} = \: - i\dot{\widetilde\psi^\dagger} - i\widetilde\psi^\dagger \underline{A} = \: & 0 \,,
\end{align}
\end{subequations}
where we introduced the fluctuation operator
\begin{equation}
    \label{eqn:fluctuation_operator_def}
    A(\mathbf x, \mathbf x'; t,t') = \: \Gamma\frac{\delta F[\psi] (\mathbf x, t)}{\delta\psi(\mathbf x', t')} \,.
\end{equation}
The solutions to the saddle-point equations $\{\psi_0, i\widetilde{\psi}_0\}$ are given by $i\widetilde{\psi}_0 = 0$ and $\dot\psi_0=\Gamma F[\psi_0]$, which constitute the \textit{deterministic} solutions of the dynamics.
In order to find the lowest order correction in fluctuations around the saddles, we decompose the field and the response field in terms of background and fluctuations
\begin{subequations}
\begin{align}
    \psi = \: & \psi_0 + \sqrt{\gamma} \psi_1 \,, \label{eqn:small_noise_field} \\
    \widetilde\psi = \: & \sqrt{\gamma} \widetilde\psi_1 \,. \label{eqn:small_noise_response_field}
\end{align}
\end{subequations}
We also re-scale the external sources $J,\widetilde{J} \mapsto J/\sqrt{\gamma}, \widetilde{J}/\sqrt{\gamma}$ for consistency \cite{chow2015path,Thomas_2014}, keeping in mind that the expression of $\underline{C}$ and $\underline{G}$ in terms of derivatives of $Z[J,\widetilde J]$ must be updated accordingly.
The expansion of the action $\mathscr{A}[\psi, \widetilde{\psi}]$ from Eq.\,\eqref{eqn:MSR_action} to quadratic order in the fluctuating corrections reads 
\begin{align}
    \frac{1}{\gamma}\mathscr{A}[\psi,\widetilde\psi] = \: &  {\frac{1}{\gamma} \mathscr{A}[\psi_0,0] - i\widetilde\psi_1^\dagger \, i\widetilde\psi_1 + i\widetilde\psi_1^\dagger \, \underline{L} \, \psi_1  + \mathcal{O}(\gamma)\,},
\end{align}
where we have introduced the operator $\underline{L}$ which implements the linearised Langevin equation, namely
\begin{align}
    L(\mathbf x, \mathbf x'; t,t') \equiv \:  \frac{\delta^2 \mathscr{A}[\psi,\widetilde\psi]}{\delta i\widetilde\psi(\mathbf x, t) \delta\psi(\mathbf x', t')}\Bigg\rvert_{\substack{\psi=\psi_0\\ \widetilde\psi=\widetilde\psi_0}}
    = \: \delta(t-t') \delta^{(d)}(\mathbf x-\mathbf x') \partial_t - A_0(\mathbf x, \mathbf x'; t,t')\,.
    \label{eqn:definition_L}
\end{align}
We have expressed the fluctuation operator in the background $\psi_0$ using the definition in Eq.\,\eqref{eqn:fluctuation_operator_def},
\begin{equation}
    \label{eqn:definition_A0}
     A_0(\mathbf x, \mathbf x';t,t') \equiv \: A(\mathbf x, \mathbf x'; t,t')\Bigg\rvert_{\substack{\psi=\psi_0}}\,.
\end{equation}
{ {
The central focus of this work is to study theories where $A_0$ is \emph{not} self-adjoint, namely
\begin{equation}
	\label{eqn:A_0_non-Hermitian_definition}
	\underline{A}_0^\dagger \neq \underline{A}_0 \,.
\end{equation}
In coordinate space, this reduces to the condition $A_0^*(\mathbf x', \mathbf x) \neq A_0(\mathbf x, \mathbf x')$. However, Eq.\,\eqref{eqn:A_0_non-Hermitian_definition} can be understood, and checked, in any basis.
}
}

 {Back to the small noise expansion}, this procedure computes perturbative corrections to the deterministic solutions of the equation of motion Eq.\,\eqref{eqn:Langevin_eq}. 
Indeed, one can check that the $\mathcal{O}(1)$ terms in the expansion of $\gamma^{-1}\mathscr{A}[\psi, \widetilde{\psi}]$ provide the dynamic description of lowest order fluctuating corrections in the Langevin approach previously used in Ref. \cite{LC21}.

The generating functional in the saddle-point expansion reads
\begin{align}
    Z[J,\widetilde J] = \:  \mathcal{N} e^{\widetilde J^\dagger\psi_0/\sqrt{\gamma}} \int \mathcal{D} [\psi_1,\widetilde\psi_1] \exp \Big (& - i\widetilde\psi_1^\dagger \, \underline{L} \, \psi_1 +  {i\widetilde\psi_1^\dagger \, i\widetilde\psi_1} \notag \\  
    &+\widetilde J^\dagger \psi_1 + J^\dagger i\widetilde\psi_1 \Big) \Big[ 1+ \mathcal{O}(\gamma) \Big] \,,
\end{align}
after truncating the expansion to quadratic order in the fluctuations.
The resulting generating functional can be evaluated directly by Gaussian integration, first in $\widetilde\psi_1$ and then in $\psi_1$. We arrive at 
\begin{equation}
    Z[J,\widetilde J] = \: \exp \left\{ \widetilde J^\dagger  \psi_0/\sqrt{\gamma} +  \widetilde J^\dagger \underline{L}^{-1} J +  {\widetilde J^\dagger \underline{L}^{-1} (\underline{L}^{-1})^\dagger \widetilde J} \right\} \,\Big[ 1+ \mathcal{O}(\gamma) \Big] \,.
\end{equation}
From it, we can directly find the correlation and response  
using Eqs. \eqref{eqn:correlation_from_Z} and~\eqref{eqn:response_from_Z}:
\begin{subequations}
\begin{align}
     \underline{C} = \: \gamma \frac{\delta^2 Z[J,\widetilde J]}{\delta \widetilde J(\mathbf x, t) \delta\widetilde J(\mathbf x', t')} \Bigg\rvert_{\substack{J=0\\\widetilde J=0}} = \: &  \psi_0 \psi_0^\dagger +  {2 \gamma \, \underline{L}^{-1} (\underline{L}^{-1})^\dagger \,,} \label{eqn:correlation_with_disconnected_component} \\
    \underline{G} = \: \hphantom{\gamma} \frac{\delta^2 Z[J,\widetilde J]}{\delta \widetilde J(\mathbf x, t) \delta \vphantom{\widetilde J} J(\mathbf x', t')} \Bigg\rvert_{\substack{J=0\\\widetilde J=0}} = \: & \underline{L}^{-1} \,. \label{eqn:response_from_MSRJD}
\end{align}
\end{subequations}
Note the Eqs. \eqref{eqn:correlation_from_Z} and~\eqref{eqn:response_from_Z} have been modified according to the re-scaling of the external sources.
The term $\psi_0\psi_0^\dagger$ is the disconnected contribution to the correlation and yields the leading order  contribution to the FD difference when $\psi_0$ depends on time.
Since we are interested in isolating the stochastic contribution to the non-equilibrium evolution of the system, from now on we always assume $\psi_0$ to be a stationary solution of the deterministic equation of motion (EoM), namely $F[\psi_0]=0$.
Then, the disconnected part of $\underline{C}$ plays no role at the level of the FDT, because only the time derivative of the correlation $\underline{C}$ enters the definition of the FD difference in Eq.\,\eqref{eqn:def_FDT_violation}.
Therefore, we will drop it in what follows and assume we have redefined the correlator $\underline{C}$ to be the connected component of the two-point function $\langle\psi\psi\rangle$.
We recall that the connected correlators are generated by the Schwinger functional
\begin{align}
    W[J, \widetilde{J}] = \log Z[J, \widetilde{J}] = \widetilde J^\dagger \psi_0/\sqrt{\gamma} +  \widetilde J^\dagger \underline{L}^{-1} J +   {\widetilde J^\dagger \underline{L}^{-1} (\underline{L}^{-1})^\dagger \widetilde J} + \mathcal{O}(\gamma),
\end{align}
from which we can recover the connected components of the two-point functions by acting with functional derivatives
\begin{subequations}
\begin{align}
    \underline{C} = \: \gamma \frac{\delta^2 W[J,\widetilde J]}{\delta \widetilde J(\mathbf x, t) \delta\widetilde J(\mathbf x', t')} \Bigg\rvert_{\substack{J=0\\\widetilde J=0}} = \: &  {2\gamma\,  \underline{L}^{-1} (\underline{L}^{-1})^\dagger \,,} \\
    \underline{G} = \: \hphantom{\gamma} \frac{\delta^2 W[J,\widetilde J]}{\delta \widetilde J(\mathbf x, t) \delta \vphantom{\widetilde J} J(\mathbf x', t')} \Bigg\rvert_{\substack{J=0\\\widetilde J=0}} = \: &  \underline{L}^{-1}\,.
\end{align}
\end{subequations}
When working at higher order in the noise amplitude $\gamma$, it is convenient to work directly with $W[J,\widetilde{J}]$ to avoid subtracting the disconnected contributions by hand.
However, this will not be relevant for our discussion.

\subsection{Uniform background}
\label{sec:FDT_viol_uniform_background}
 {So far, we have  {performed} the small noise expansion in the response-field formalism and arrived at Eqs.\,\eqref{eqn:correlation_with_disconnected_component} and~\eqref{eqn:response_from_MSRJD}, which are valid for any background $\psi_0$. 
In this subsection, we focus now on \emph{uniform} backgrounds, \textit{i.e.} $\psi_0$ does not depend on either time or space coordinates.   {In section~\ref{sec:langevin}} we will also extract results for more general inhomogeneous stationary backgrounds.
Since $\psi_0$ is a constant, both time- and space-translational invariance are preserved.
 {Thus, the response and correlation functions only depend on the relative coordinates $\mathbf x- \mathbf x'$ and $t-t'$,}
and it is convenient to work in frequency and momentum space.
We define
\begin{align}
    G(\mathbf{x},\mathbf{x'}; t,t') = \: & \int \!\!\frac{d\omega}{2\pi} \int\!\!\frac{d^d \textbf{q}}{(2\pi)^d} e^{-i\omega (t-t') - i\mathbf q \cdot(\mathbf x-\mathbf x')} \widehat G (\omega,\mathbf q) \,,
\end{align}
with $\widehat{G}$ the Fourier transform of $\underline{G}$ (both in space and time), and analogously for the correlation $\underline{C}$ and the FD difference $\underline{\Delta}$.
 {Since we are working in a uniform background,} the linearised Langevin operator in Fourier space is diagonal and reads 
\begin{equation}
    \widehat L(\omega,\mathbf q) = \: -i\omega { {-}} \lambda_{\mathbf q}\,,
\end{equation}
where $\lambda_{\mathbf q}$ are the eigenvalues of the fluctuation operator $\underline{\widehat A}_0$ in momentum space. Importantly, we allow for \emph{complex} eigenvalues, as we aim to account for the possibility that the
 {fluctuation operator $\underline{A}_0$} is non-Hermitian.
  Given that the operator $\underline{A}_0$ in position space is real, its eigenvalues satisfy the following property
\begin{equation}
    \label{eqn:parity_breaking_non_hermiticity_equivalence}
    \lambda_{-\mathbf q} = \lambda_{\mathbf q}^* \, \Longleftrightarrow \, \ \mathrm{Re}\, \lambda_{-\mathbf q} = \mathrm{Re} \, \lambda_{\mathbf q} \,, \ \mathrm{Im} \, \lambda_{-\mathbf q} = - \mathrm{Im} \, \lambda_{\mathbf q} \,.
\end{equation}
{This validates our intuition that, when  {linearising} about a uniform background, non-Hermiticity and parity-breaking are equivalent.
 {Let us be even more explicit. Say we start by being completely agnostic about the hermiticity properties of the linear theory. Then, Eq.\,\eqref{eqn:parity_breaking_non_hermiticity_equivalence} implies that, if the fluctuation operator $\underline{A}_0$ is parity odd, which in Fourier space means $\lambda_\mathbf{q}\neq\lambda_{-\mathbf{q}}$, the eigenvalues must possess an imaginary part, namely $\underline{A}_0$ is non-Hermitian. 
Conversely, if we assume $\underline{A}_0$ is non-Hermitian, then $\mathrm{Im}\lambda_\mathbf{q}\neq0$ and $\lambda_\mathbf{q}\neq\lambda_{-\mathbf{q}}$.
Crucially, for this argument to hold, the fluctuation operator must admit a diagonal representation in Fourier space, which is only true when the theory is linearised about a uniform background. As we shall see, the equivalence of parity breaking and non-Hermiticity of the fluctuation operator fails when linearising about a non-uniform background, as shown in section~\ref{sec:domain_wall}.}

Intuitively, parity-breaking may arise at the linear level when the linearisation yields terms of odd parity, being thus related to violation of spatial inversion symmetries. This is exemplified in section \ref{sec:langevin_homogeneous_NR_A}, motivated by the continuous description of Ising spins anisotropically interacting through vision--cones \cite{Garc_s_2025, akritidis2026fate} which break rotational symmetry.} 
In momentum space, the response and the correlator take on a simple form 
\begin{subequations}
\begin{align}
    \widehat G(\omega,\mathbf q) = \: & \frac{1}{-i\omega - \lambda_{\mathbf q}} \,, \\
    \widehat C(\omega, \mathbf q) = \: & 2\gamma \, \left| \widehat G (\omega, \mathbf q) \right|^2 \,,
\end{align}
\end{subequations}
and the FD difference can be recast to
\begin{align}
     \widehat \Delta(\omega,\mathbf q) = \: & \widehat G(\omega,\mathbf q) - \widehat G(-\omega,\mathbf q) - \gamma^{-1}i\omega \, \widehat C(\omega,\mathbf q)
    = \: - \frac{4\omega}{\omega^2 + \lambda_{\mathbf q}^2} \frac{\mathrm{Im} \, \lambda_{\mathbf q}}{i\omega - \lambda_{\mathbf q}^*} \,.
    \label{eqn:FDT_violation_momentum}
\end{align}
It immediately follows that the FDT is violated if and only if the eigenvalues $\lambda_{\mathbf q}$ have a non-vanishing imaginary part.
 {Once again,} this   {is a manifestation of} the response $\widehat G(\omega, \textbf{q})$ \emph{not} being parity even, namely
\begin{equation}
  \Delta\neq 0  \quad \Longleftrightarrow \quad \mathrm{Im} \, \lambda_{\mathbf q} \neq 0 \quad \Longleftrightarrow  \quad \widehat G (\omega, \mathbf q) \neq \widehat G(\omega, -\mathbf q) \,.
\end{equation}
We can define $\epsilon_{\mathbf q}$ and $\zeta_{\mathbf q}$ as the real and imaginary parts of the eigenvalue $\lambda_{\mathbf q}$ and expand for small $\zeta_{\mathbf q}$ to find the leading order behaviour for small non-Hermiticity.
We find
\begin{equation}
    \label{eqn:FDT_violation_small_zeta}
    \widehat \Delta(\omega,\mathbf q) = \: \frac{4\omega}{\omega^2 + \epsilon_{\mathbf q}^2} \left[ \frac{\zeta_{\mathbf q}}{-i\omega + \epsilon_{\mathbf q}} - \frac{i\zeta_{\mathbf q}^2}{\omega^2 + \epsilon_{\mathbf q}^2} + \mathcal{O}(\zeta_{\mathbf q}^3)\right] \,,
\end{equation}
 {to order $\mathcal{O}(\gamma^0)$}. To summarise, we find that {a complex spectrum of the fluctuation operator $\underline{A}_0$ leads to non-equilibrium dynamics and violations of the FDT; non-Hermitian terms in $\underline{A}_0$ are required for it.}
This can be quantified for a uniform stationary state using Eq.\,\eqref{eqn:FDT_violation_momentum}, and the FD difference is linear in the non-Hermiticity $\zeta_{\mathbf q}$ at lowest order, as in Eq.\,\eqref{eqn:FDT_violation_small_zeta}.

The violation of the FDT is a hallmark of non-equilibrium, and it can be extracted from simulations and experiments by measuring the correlation and the response functions directly.
On the other hand, a convenient thermodynamic and informatic measure of the non-equilibrium dynamics of a system is the entropy production.
In the following, we express the entropy production in terms of the FD difference via a Harada--Sasa relation, and we use Eq.\,\eqref{eqn:FDT_violation_momentum} to compute it in the limit of small non-Hermiticity.

\subsection{Harada--Sasa relation for entropy production}
\label{sec:harada-sasa}
The entropy production is an informatic measure of the time irreversibility of a stochastic dynamics \cite{CN17,LC21}. 
It is defined by the ratio between probabilities of trajectories forward and backwards in time.
Within the MSRJD path-integral formalism, the probability of a trajectory in the field-space $\{\psi,\widetilde\psi\}$ is given by
\begin{equation}
    \mathcal{P}[\psi,\widetilde\psi] \propto \: e^{-\mathscr{A}[\psi,\widetilde\psi]}\,,
\end{equation}
up to normalisation.
The entropy production  {within the response-field formalism} then follows naturally 
\begin{align}
    S[\psi] \equiv \:   \log \frac{\mathcal{P}[\psi,\widetilde\psi]}{\mathcal{P}[\mathcal{T}\psi,\mathcal{T}\widetilde\psi]} 
    &= \: \mathscr{A}[\mathcal{T}\psi,\mathcal{T}\widetilde\psi] - \mathscr{A}[\psi,\widetilde\psi] \notag \\
    &= \:  \beta \int_{\textbf{x},t}  \dot\psi (\vec x, t) \, F[\psi] (\vec x, t)\,, 
    \label{eqn:definition_entropy_production}
\end{align}
directly from Eq.\,\eqref{eq:broken_invariance_action}. Alternatively, one can use the canonical form of $\mathscr{A}[\psi, \widetilde{\psi}]$ in Eq.\,\eqref{eqn:generating_functional_canonical} and correspondingly define $S[\psi]$ by adapting the definition of the $\mathcal{T}$-operation. The resulting expression for $S[\psi]$ is the same.

As we anticipated, the entropy production does not depend on the response field $\widetilde\psi$, which can be integrated out.
This implies one can equivalently work with the Onsager-Machlup functional in all instances where it is well defined \cite{cugliandolo2017rules}.
Rather than the global entropy production, it is convenient to study the average \textit{local entropy production rate} (LEPR) $\sigma_\psi$, especially useful for non-uniform backgrounds.
It is defined by the relation
\begin{equation}\label{eq:loc_decomposition_epr}
    \langle S[\psi] \rangle = \: \int_{\textbf{x},t} \: \sigma_\psi(\vec x, t) \,,
\end{equation}
and, using Eq.\,\eqref{eqn:definition_entropy_production}, 
\begin{equation}
    \sigma_\psi(\vec x, t) = \: \beta \big\langle \dot\psi (\vec x, t) \, F[\psi] (\vec x, t) \big\rangle\,.
\label{eqn:definition_LEPR}
\end{equation}
Our goal is to obtain a Harada--Sasa type of relation, namely an equation that expresses the average LEPR $\sigma_\psi$, and consequently the integrated entropy production $\langle S[\psi]\rangle$, in terms of the FD difference $\underline{\Delta}$ \cite{Harada05}. 
To achieve this, consider the identity\,\cite{zinn-justin}
\begin{align}
    0 = & \int \mathcal{D}[\psi,\widetilde\psi] \frac{\delta}{\delta i\widetilde\psi(\vec x', t')} \left( \mathcal{O}[\psi,\widetilde\psi]  e^{- \mathscr{A}[\psi,\widetilde\psi]} \right) 
    = \left\langle \frac{\delta \mathcal{O}[\psi,\widetilde\psi]}{\delta i\widetilde\psi(\vec x',t')} \right\rangle -  \left\langle \mathcal{O}[\psi,\widetilde\psi] \frac{\delta \mathscr{A}[\psi,\widetilde\psi]}{\delta i\widetilde\psi(\vec x',t')} \right\rangle \,,
\end{align}
for an arbitrary observable $\mathcal{O}[\psi, \widetilde{\psi}]$, which holds because the probability density $e^{-\mathscr{A}[\psi,\widetilde\psi]}$ decays fast enough so as not to generate boundary terms.
Then, choosing $\mathcal{O}[\psi,\widetilde\psi]= \psi(\vec x,t)$, we find
\begin{align}
    0 = \: & \left\langle \psi(\vec x, t) \left( \dot\psi(\vec x',t') - \Gamma F[\psi](\vec x', t') - 2\gamma i \widetilde\psi(\vec x', t') \right) \right\rangle \notag \\[1.5ex]
    = \: &  \, \partial_{t'} C(\vec x, \vec x'; t,t') - 2\gamma \, G(\vec x, \vec x'; t,t')
    - \big \langle \psi(\vec x, t) \Gamma F[\psi](\vec x', t')\big \rangle \,.
\end{align}
Note that we are working with the action $\mathscr{A}$ as defined in Eq\, \eqref{eqn:MSR_action}, before rescaling the response field $\widetilde{\psi}$. Now, dividing by $2\gamma$ and antisymmetrising the right-hand-side by $t\leftrightarrow t'$, we find the following identities
\begin{subequations}
    \begin{align}
    G(\textbf{x},\textbf{x}'; t,t') - \frac{1}{2}\gamma^{-1}\partial_{t'}C(\vec x, \vec x'; t,t') &= -\frac{1}{2}\,\beta \, \big \langle \psi(\vec x, t) F[\psi](\vec x', t') \big \rangle, \label{eqn:first_identity_hs}\\ 
    G(\vec x, \vec x'; t',t) - \frac{1}{2}\gamma^{-1}\partial_t C(\vec x, \vec x';t',t) & = - \frac{1}{2} \,\beta \big \langle \psi(\vec x, t')F[\psi](\vec x', t)\big \rangle.
\end{align}
\end{subequations}
Taking the space-diagonal part of these identities and subtracting them, we obtain 
\begin{align}
     \Delta (\vec x, \vec x; t,t') = \frac{\beta}{2}\big \langle \psi(\vec x, t')F[\psi](\vec x, t) - \psi(\vec x, t)F[\psi](\vec x, t') \big \rangle.
\end{align}
Now, let us act with a derivative with respect to $t$ on both sides of this equation. 
At this point, we recall that in any stationary background, two-point functions are time-translation invariant and only depend on $t-t'$. 
Therefore, acting with $\partial_t$ is equivalent to acting with $-\partial_{t'}$, so that
\begin{align}
    \frac{\beta}{2} \partial_t \big \langle \psi(\vec x, t')F[\psi](\vec x, t) - \psi(\vec x, t)F[\psi](\vec x, t') \big \rangle = \: &
    - \frac{\beta}{2} \big \langle \dot\psi(\vec x, t')F[\psi](\vec x, t) + \dot\psi(\vec x, t)F[\psi](\vec x, t') \big \rangle \notag \\
    \xrightarrow{t'\to t} \: & - \beta \big \langle \dot \psi(\vec x, t)F[\psi](\vec x, t) \big \rangle \notag \\
    \overset{\eqref{eqn:definition_LEPR}}{=} \: & -\sigma_\psi (\vec x) \,.
\end{align}
 {Note that $\sigma_\psi$ \emph{does not depend on time}, because equal-time averages in a stationary background are time-independent.}
Finally, we get the following Harada--Sasa relation that expresses the average LEPR in terms of the FD difference,
\begin{align}
    \sigma_\psi(\vec x) = \: - \, \mathrm{diag}_{\vec x,t} \, \partial_t \Delta (\vec x, \vec x'; t,t') \,,
    \label{eqn:harada_sasa_local}
\end{align}
where the diagonal of a quadratic form is $\mathrm{diag}_{\vec x,t} Q(\vec x, \vec x'; t,t') = Q(\vec x, \vec x; t,t)$.
The total entropy production is then recovered by integrating over space and time coordinates, via Eq.\,\eqref{eq:loc_decomposition_epr}, and reads
\begin{align}
    \langle S[\psi] \rangle = - \textup{tr}\, \partial_t \Delta (\vec x, \vec x'; t,t'). \label{eqn:harada_sasa_integrated}
\end{align}
 {Since $\sigma_\psi$ is time-independent, the integral over $t$ in the trace simply measures the time length of the considered trajectory.}
 {
The Harada-Sasa relation in its local and integrated form, Eqs.\,\eqref{eqn:harada_sasa_local} and~\eqref{eqn:harada_sasa_integrated}, respectively, is valid for any stationary field configuration $\psi$ and crucially does not rely on the small-noise expansion.  {Integrated Harada-Sasa relations have been derived for field theories with space- and time-translational invariance \cite{CN17, johnsrud2025}. However, the local relation Eq.~(\ref{eqn:harada_sasa_local}) requires time-translational invariance only, and therefore applies to spatially inhomogeneous systems.}}

 {
Having established the local Harada--Sasa relation\,\eqref{eqn:harada_sasa_local} for arbitrary stationary backgrounds, it is a quick exercise to apply it for a uniform background as studied in section~\ref{sec:FDT_viol_uniform_background}.
We have already extracted the FD difference in momentum space and arrived at Eq.\,\eqref{eqn:FDT_violation_momentum}.
Then,}
 the average LEPR is quickly computed by plugging Eq.\,\eqref{eqn:FDT_violation_momentum} into Eq.\,\eqref{eqn:harada_sasa_local}, and it amounts to
\begin{align}
    \label{eq:epr_from_fd}
    \sigma_\psi = \int_{\omega, \textbf{q}}\, i\omega \widehat{\Delta}(\omega, \vec q) = \int_{\omega, \vec q} \, \frac{4 \omega^2 \zeta_{\vec q}^2}{| \omega^2 + \lambda_{\vec q}^2|^2} = \int_{\vec q} \, \frac{\zeta_{\vec q}^2}{|\epsilon_{\vec q}|}
\end{align}
 {to order $\mathcal{O}(\gamma^0)$}. In the second step, we have used that only the even part of the integrand under $\omega\to-\omega$ and $\mathbf q\to -\mathbf q$ survives.
We learn that the average LEPR is quadratic in $\zeta_{\mathbf q}$, as opposed to the FD difference $\underline{\Delta}$, which is non-zero at linear order.
As a consequence, when mirroring the reference frame so that $\vec q\to-\vec q$, the FD difference changes sign, while the average LEPR remains positive.
This agrees with our intuition that the average entropy must always and everywhere increase, regardless of the frame of reference. 
We highlight that this does not prevent individual realisations from proceeding along a decreasing entropy trajectory.
As a last note, we stress that Eq.\,\eqref{eq:epr_from_fd} is the leading order in a small-noise expansion and should be trusted only in the limit of small noise amplitude $\gamma$, under the assumptions of white noise and uniform backgrounds.
 {In the rest of this article, we discuss an alternative route to extract the average LEPR in the small noise expansion. As we will see, this will allow us to provide results even for inhomogeneous stationary backgrounds.}

\section{Entropy Production rate from the Langevin equation}
\label{sec:langevin}

In this section, we follow the approach developed by Li and Cates in Ref.\,\cite{LC21} to extract the LEPR of the system directly from the Langevin description.
The advantage of this approach is that we find a general formula for the LEPR that is valid for any stationary state, including non-uniform ones.
We demonstrate the power of this formula by computing the LEPR analytically around a phase boundary due to a non-Hermitian interaction term.

\subsection{Local EPR as a generic mode expansion}

We start from the Langevin equation~\eqref{eqn:Langevin_eq} and split the field~$\psi$ in a stationary background~$\psi_0$ plus small fluctuations~$\psi_1$, as in Eq.\,\eqref{eqn:small_noise_field}.
 {We also allow the noise statistics to have non-trivial correlations in space by the introduction of a noise kernel $\underline{K}$, since the framework in Ref. \cite{LC21} can be carried altogether.} We find the linearised Langevin equation for the fluctuations, which reads
\begin{equation}
    \dot\psi_1 = \: \underline{A}_0 \psi_1 + \sqrt{2} \xi \,,
\end{equation}
where the fluctuation operator $\underline{A}_0$ in the stationary background $\psi_0$ has been defined in Eq.\,\eqref{eqn:definition_A0}.
Equivalently, we could write $\underline{L} \psi_1 = \sqrt{2} \xi$, with the linearised Langevin operator $\underline{L}$ defined in Eq.\eqref{eqn:definition_L}.
In a stationary background, the time dependence of the fluctuation operator is trivial, and for our purposes, it is convenient to factor it out by defining the operator $\underline{\mathcal{A}}_0$ via the relation
\begin{equation}
    A_0(\vec x, \vec x'; t,t') = \: \delta(t-t') \mathcal{A}_0(\vec x, \vec x') \,.
\end{equation}
Next, we introduce the covariance matrix $\underline{\mathcal{C}}$, as the equal time limit of the fluctuation correlation function, namely,
\begin{equation}
    \label{eqn:covariant_matrix_def}
    \gamma\, \mathcal{C} (\mathbf x, \mathbf x') \equiv \: \gamma \langle \psi_1(\mathbf x, t) \psi_1(\mathbf x', t) \rangle = \: C(\mathbf x, \mathbf x'; t,t) \,,
\end{equation}
where we take out a factor $\gamma$ so that $\underline{\mathcal{C}}$ is of order $\mathcal{O}(1)$.
We further assume that the noise is $\delta$-correlated in time, so that we can write
\begin{equation}
    \label{eqn:time_delta_correlated_noise}
    \langle \xi(\mathbf x, t)\, \xi(\mathbf x', t') \rangle = \: \delta(t-t') \mathcal{K} (\mathbf x, \mathbf x') \,,
\end{equation}
where $\underline{\mathcal{K}}$ is the spatial component of the noise kernel.
For these space-dependent quantities, we can use the same matrix notation as in the previous section, dropping the time integration from the product.

Following Ref.\,\cite{LC21}, the average LEPR can be written fully in terms of the fluctuation operator, the covariance matrix, and the noise kernel, and it reads
\begin{equation}
    \label{eqn:average_LEPR}
    \sigma_\psi (\mathbf x) = \: \mathrm{diag}_{\mathbf x} \left[ \kappa^{-1} \underline{\mathcal{F}}_a \,\underline{\mathcal{C}} \,\underline{\mathcal{F}}_a^\dagger (\kappa^{-1})^\dagger \right] \,,
\end{equation}
where $^\dagger$ stands for Hermitian conjugation. The factor $\kappa$ is a ``square root'' of the spatial noise kernel, that is, $\kappa\kappa^\dagger = \underline{\mathcal{K}}$, and it is sometimes referred to as Choleski factor. 
We have introduced the ``anti-symmetric'' part of the dynamics
\begin{equation}
    \underline{\mathcal{F}}_a = \: \underline{\mathcal{A}}_0 { {+}} \underline{\mathcal{K}}\,\underline{\mathcal{C}}^{-1} \,.
\end{equation}
Equation~\eqref{eqn:average_LEPR} has been first obtained by Li and Cates in Ref.\,\cite{LC21}, where the interested reader can find a detailed derivation.
They relied on the following set of assumptions.
First, they work with a single real scalar field that obeys the Langevin dynamics in Eq.\,\eqref{eqn:Langevin_eq}.
Second, the noise is delta-correlated in time as in Eq.\,\eqref{eqn:time_delta_correlated_noise}.
Finally, they study the dynamics in the small-noise expansion about a stationary configuration, as defined in Eq.\,\eqref{eqn:small_noise_field}.
Then, the entropy production can be written as a sum of an \textit{external} contribution, that is due to the external force encoded in $\underline{\mathcal{A}}_0$, and an \textit{internal contribution}, which relates to how the probability density of the configuration changes over time along the field trajectory and is here represented by $\underline{\mathcal{K}}\,\underline{\mathcal{C}}^{-1}$.

By applying their formula, Li and Cates conclude that the average LEPR vanishes at lowest order in $\gamma$ if the noise kernel $\underline{\mathcal{K}}$ and the fluctuation operator $\underline{\mathcal{A}}_0$ commute. In particular, this requires the noise kernel to be non-trivial.
As we shall see, this is only the case when the fluctuation operator is Hermitian.
We show this next by deriving a closed-form expression for the average LEPR in the presence of a small non-Hermitian term.

To compute the average LEPR using Eq.\,\eqref{eqn:average_LEPR}, we require the covariance matrix $\mathcal{C}$,
which can be found by solving the Lyapunov equation\,\cite{gardiner2010stochastic,van2011stochastic}
\begin{equation}
    \label{eqn:Lyapunov_eq_full}
    \underline{\mathcal{A}}_0 \, \underline{\mathcal{C}} + \underline{\mathcal{C}} \,\underline{\mathcal{A}}_0^\dagger = \: { {-}} 2 \underline{\mathcal{K}} \,,
\end{equation}
We split the fluctuation operator into a Hermitian and an anti-Hermitian part
\begin{equation}
    \underline{\mathcal{A}}_0 = \: \underline{\mathcal{A}}_0^{\mathrm{H}} + \underline{\mathcal{A}}_0^{\mathrm{AH}} \,,
\end{equation}
where $(\underline{\mathcal{A}}_0^{\mathrm{H}} )^\dagger=\underline{\mathcal{A}}_0^{\mathrm{H}}$ and $(\underline{\mathcal{A}}_0^{\mathrm{AH}})^\dagger = - \underline{\mathcal{A}}_0^{\mathrm{AH}}$.
Note that here we depart from the working hypothesis of Li and Cates, who assume that the fluctuation operator is Hermitian and that its eigenvalues are real.
Under these assumptions, they obtain a non-vanishing average LEPR at leading order in $\gamma$ only if $\underline{\mathcal{A}}_0$ and $\underline{\mathcal{K}}$ do not commute.
To isolate the contribution of the non-Hermitian part, let us assume that $\underline{\mathcal{K}}$ and $\underline{\mathcal{A}}_0^{\mathrm{H}}$ \textit{do} commute instead.
Then, at leading order in the limit of small anti-Hermitian part $\underline{\mathcal{A}}_0^{\mathrm{AH}}$, Eq.\,\eqref{eqn:Lyapunov_eq_full} is solved by 
\begin{equation}
    \label{eqn:covariance_matrix_small_ah}
    \underline{\mathcal{C}} = \: { {-}}\left(\underline{\mathcal{A}}_0^{\mathrm{H}}\right)^{-1} \underline{\mathcal{K}} \,.
\end{equation}
We can verify that this is a good solution by plugging it into Eq.\,\eqref{eqn:Lyapunov_eq_full} and dropping the anti-Hermitian part.
One finds $\underline{\mathcal{A}}_0^{\mathrm{H}} \, \underline{\mathcal{C}} + \underline{\mathcal{C}} \,\underline{\mathcal{A}}_0^{\mathrm{H}} = - 2\underline{\mathcal{K}} - (\underline{\mathcal{A}}_0^{\mathrm{H}})^{-1}[\underline{\mathcal{K}},\underline{\mathcal{A}}_0^{\mathrm{H}}]$.
We see now, that the ansatz \eqref{eqn:covariance_matrix_small_ah} is indeed a good solution if $\underline{\mathcal{K}}$ and $\underline{\mathcal{A}}_0^{\mathrm{H}}$ commute, i.e. if $[\underline{\mathcal{K}},\underline{\mathcal{A}}_0^{\mathrm{H}}] = \underline{\mathcal{K}}\,\underline{\mathcal{A}}_0^{\mathrm{H}} - \underline{\mathcal{A}}_0^{\mathrm{H}}\,\underline{\mathcal{K}} = 0$.

We could be tempted to make a similar ansatz for the solution to the full Lyapunov equation by writing $\underline{\mathcal{C}}=-\underline{\mathcal{A}}_0^{-1}\underline{\mathcal{K}}$. Although, strictly speaking, this solves Eq.\,\eqref{eqn:Lyapunov_eq_full}, the resulting covariance matrix is \emph{not} Hermitian, unless $\underline{\mathcal{A}}_0$ is.
However, by the definition~\eqref{eqn:covariant_matrix_def}, the covariance matrix must be Hermitian, meaning that the said ansatz does not fulfil the required properties.

At this point, we are ready to compute the average LEPR. We take the covariance matrix as obtained in Eq.\,\eqref{eqn:covariance_matrix_small_ah} and plug it into Eq.\,\eqref{eqn:average_LEPR}.
The anti-symmetric part of the dynamics now takes a more suggestive form
\begin{equation}
    \mathcal{F}_a = \: \underline{\mathcal{A}}_0 { {+}} \underline{\mathcal{K}}\,\underline{\mathcal{C}}^{-1} = \: \underline{\mathcal{A}}_0^{\mathrm{AH}}\,,
\end{equation}
and the average LEPR reads
\begin{equation}
    \label{eqn:LEPR_small_ah_generic}
    \sigma_\psi (\vec x) = \: \mathrm{diag}_{\mathbf x} \left[ \kappa^{-1} \underline{\mathcal{A}}_0^{\mathrm{AH}} \underline{\mathcal{C}} \left(\underline{\mathcal{A}}_0^{\mathrm{AH}}\right)^\dagger (\kappa^{-1})^\dagger\right] \,.
\end{equation}
Thus, we confirm our result from the previous section and find that the average LEPR is quadratic in the non-Hermiticity of the  {fluctuation operator, at leading order in both non-Hermiticity and noise amplitude}.
Furthermore, Eq.\,\eqref{eqn:LEPR_small_ah_generic} extends the previous result by Li and Cates in Ref.\,\cite{LC21} to non-Hermitian systems.
In particular, we show that, even when the noise kernel and the fluctuation operator are simultaneously diagonalisable, the average LEPR does not need to vanish.

In the rest of this section, we apply Eq.\,\eqref{eqn:LEPR_small_ah_generic} to the case of both a uniform and a non-uniform stationary state in the non-reciprocal Model A. For simplicity, we work with Gaussian white noise, namely $\underline{\mathcal{K}}=\mathds{1}$, and the function $\kappa$ can be chosen to be a unit vector in the function space.

\subsection{Local EPR of a uniform state}
At the end of section~\ref{sec:harada-sasa},  {we applied the Harada--Sasa relation\,\eqref{eqn:harada_sasa_local} to compute}  the average local EPR in a uniform state, and arrived at Eq.\,\eqref{eq:epr_from_fd}.
Here, we start from the expression of the local EPR  Eq.\,\eqref{eqn:LEPR_small_ah_generic} to show that, under the same assumptions, we recover the same expression.

In the uniform phase, we can work in momentum space, where both the fluctuation operator and the covariance matrix become diagonal.
Following the notation introduced in section~\ref{sec:FDT_viol_uniform_background}, the fluctuation operator reads 
\begin{equation}
    \mathcal{A}_0(\mathbf x, \mathbf x') = \: \int \frac{d^d \vec q}{(2\pi)^d} e^{-i\mathbf q\cdot (\mathbf x-\mathbf x')} \left( \epsilon_{\mathbf q} + i \zeta_{\mathbf q} \right) \,,
\end{equation}
where $\epsilon_{\mathbf q}$ and $\zeta_{\mathbf q}$ are both real.
Recall that we are working in the limit of small non-Hermiticity $\zeta_{\mathbf q}$.
The covariance matrix can be promptly found to be
\begin{equation}
    \mathcal{C}(\mathbf{x}, \mathbf{x}') = \: { {-}} \int \frac{d^d \vec q}{(2\pi)^d} e^{-i\mathbf q\cdot (\mathbf x-\mathbf x')} \frac{1}{\epsilon_{\mathbf q}} \,.
\end{equation}
Altogether, we can compute the average LEPR in the uniform phase using Eq.\,\eqref{eqn:LEPR_small_ah_generic} and find
\begin{equation}
    \label{eqn:average_LEPR_Fourier}
     {\sigma_\psi = \:  \int_{\textbf{q}}  \frac{\zeta_{\mathbf{q}}^2}{|\epsilon_{\mathbf q}|}}
\end{equation}
 {at leading order in noise amplitude}, which matches Eq.\,\eqref{eq:epr_from_fd} exactly.

\subsection{Application to the Non-Reciprocal Model A}\label{sec:langevin_homogeneous_NR_A}
We illustrate now the framework using an example when Hermiticity is broken explicitly, although we recall that the formalism is equally valid when the non-Hermiticity arises spontaneously after linearising the theory around a non-uniform background.
We focus on the Non-Reciprocal Model A, a non-Hermitian extension of a $\psi^4$-theory 
first introduced in Ref.\,\cite{Garc_s_2025} as a coarse-grained description of an Ising model with vision cone interactions.  {Namely, $\mathbf{v}$ is a fixed direction that provides Ising spins on a lattice with a preferred \textit{interaction} direction, breaking reciprocity at the microscopic scale, the details of the microscopic model can be found in Ref.~\cite{Garc_s_2025}}. 
The deterministic part of its governing Langevin equation reads
\begin{equation}
    \Gamma F[\psi] = \: -r\psi + \mu \nabla^2\psi - u \psi^3 + 2\lambda \psi (\mathbf v\cdot\nabla) \psi \,,
\end{equation}
where $r$ is the reduced temperature, $\mu$ is the elasticity constant, and $u$ and $\lambda$ are the quartic and the non-reciprocal couplings, respectively.
The non-reciprocal term depends on the vision cone vector $\mathbf v$, set by the microscopic details of the system.
At high temperatures, $r$ is positive, and the only uniform stationary solution of the deterministic EoM is the ground state $\psi_0=0$. 
As the temperature lowers, $r$ becomes negative, and the system undergoes a continuous phase transition from a disordered to an ordered phase.
The field acquires a non-vanishing expectation value, and there are two stable uniform solutions to the deterministic EoM, namely $\psi_0 = \pm m = \pm \sqrt{-r/u}$, where $m$ is the magnetisation.
The fluctuation operator reads  
\begin{equation}
    \mathcal{A}_0 (\mathbf x,\mathbf x') = \: { {-}}\delta^{(d)}(\mathbf x-\mathbf x') \left[ - \mu\nabla^2 + r + 3u\psi_0^2 - 2\lambda \psi_0 (\mathbf v\cdot\nabla) \right] \,,
\end{equation}
which in Fourier space becomes
\begin{equation}
    \label{eqn:fluctuation_operator_Fourier_homogeneous}
    \widehat{\mathcal{A}}_0 (\mathbf q) = \: { {-}} \left[\mu\mathbf{q}^2 + r + 3u\psi_0^2 + 2i\lambda \psi_0 q_{\mathrm{v}}\right] \,.
\end{equation}
Here, we have defined the component of the momentum in the direction of the vision cone vector, namely $q_{\mathrm{v}} = \mathbf{q}\cdot\mathbf v$.
From Eq.\,\eqref{eqn:fluctuation_operator_Fourier_homogeneous} we can readily identify the real and imaginary parts of the eigenvalues for the fluctuation operator 
\begin{subequations}
\begin{align}
    \epsilon_{\mathbf q} = \: & { {-}} \left[ \mu\mathbf{q}^2 + r + 3u\psi_0^2 \right] \,, \\
    \zeta_{\mathbf q} = \: & - 2\lambda \psi_0 q_{\mathrm{v}} \,.
\end{align}
\end{subequations}
Using Eq.\,\eqref{eqn:average_LEPR_Fourier}, we find the average LEPR in the uniform $\psi_0$ phase to be
\begin{equation}
    \label{eqn:average_LEPR_hom}
    \sigma_\psi = \: \int_{\mathbf q} \frac{4\lambda^2\psi_0^2 q_{\mathrm{v}}^2}{\mu\mathbf{q}^2 + r + 3u\psi_0^2} + \mathcal{O}(\gamma)\,.
\end{equation}
A few immediate observations about our result are in order.
First of all, the LEPR is explicitly non-negative, in accordance with the second law of thermodynamics.
Specifically, it is proportional to the square of the non-reciprocal coupling $\lambda$, meaning that in the absence of non-reciprocity the system is in equilibrium.
Just as observed in Ref.~\cite{SuchanekPRE2023} for the non-reciprocal Cahn-Hilliard model, the LEPR is formally divergent since the integrand converges to a finite value for $q\to\infty$, implying the need for a UV regulator.
Finally, the LEPR  {vanishes in the disordered phase $\psi_0 = 0$},  {at leading order in noise amplitude}.

Interestingly, our formula~\eqref{eqn:LEPR_small_ah_generic} is not restricted to the uniform phase, but can be used to compute the LEPR around any stationary solution of the deterministic part of the dynamics. In the next section, we use it to compute the LEPR for a domain wall background in the non-reciprocal Model A.

\subsection{Local EPR across a domain wall}
\label{sec:domain_wall}
When the system is below the critical temperature, it exhibits additional stationary solutions: domain walls.
A domain wall is an non-uniform field configuration that depends on only one spatial coordinate.
Let us call this coordinate~$x_1$, which we can always do by a rotation of the reference frame. This may or may not be parallel to the vision cone vector~$\mathbf v$. 
Then, we define~$m^2=-r/u$,~$2\alpha^2=-r/\mu$ and~$\delta=2\lambda\sqrt{2/(\mu u)}$, and the rescaled field~$f(\alpha x_1) = \psi_0(x_1)/m$ so that the deterministic part of the stationary equation becomes
\begin{equation}
    \label{eqn:rescaled_deterministic_eom}
    f''(\xi) + 2 f(\xi) \left( 1- f(\xi)^2\right) + \mathrm{v}_1 \delta f(\xi) f'(\xi) = \: 0 \,,
\end{equation}
where~$\mathrm{v}_1$ is the~$x_1$ component of vision cone vector~$\mathbf v$ in this reference frame and is zero if~$\mathbf v$ is orthogonal to the~$x_1$ axis (i.e. parallel to the domain wall).
Equation~\eqref{eqn:rescaled_deterministic_eom} has solutions~$f(\xi)=\pm1$, corresponding to the two degenerate ordered states, but also exhibits non-constant solutions.
For the time being, let us assume that the vision cone vector is parallel to the domain wall, namely $\mathrm{v}_1=0$. 
We can drop the last term in Eq.\,\eqref{eqn:rescaled_deterministic_eom}, which then becomes the famous kink equation, whose solution reads
\begin{equation}
    \label{eqn:kink}
    f_{\mathrm{K/AK}} (\xi) = \: \pm \tanh (\xi-\xi_0) \,,
\end{equation}
where~$\xi_0$ is an arbitrary integration constant that can be identified as the location of the domain wall and can be set to zero by translating the reference frame. 
The labels K and AK refer to kink and anti-kink, respectively.
The fluctuation operator in the kink background can be split into a Hermitian and an anti-Hermitian part, which read
\begin{subequations}
\begin{align}
    \mathcal{A}_{\mathrm{K}}^{\mathrm{H}} (\mathbf x, \mathbf x')= \: & { {-}} \delta^{(d)}(\mathbf x-\mathbf x') \left[ -\mu \nabla^2 - 2\mu\alpha^2 (1-3\tanh^2\alpha x_1) \right] \,, \\
    \mathcal{A}_{\mathrm{K}}^{\mathrm{AH}} (\mathbf x, \mathbf x') = \: & \delta^{(d)}(\mathbf x-\mathbf x') \; 2\lambda m \tanh \alpha x_1\, \partial_{\vec{v}} \,, \label{eqn:anti_hermitian_part_DW}
\end{align}
\end{subequations}
where we have defined $\partial_{\vec{v}}= \mathbf{v}\cdot \nabla$, the derivative in the direction of the vision cone vector.
 {We observe that the anti-Hermitian part Eq.\,\eqref{eqn:anti_hermitian_part_DW} is parity even. As we anticipated, this provides a clear example of a case where non-Hermiticity and parity-breaking are \emph{not} equivalent. Crucially, this is possible because we are expanding the theory about a domain wall, a stationary state which breaks translational invariance.
In the following, we compute a non-vanishing local EPR due to this anti-Hermitian part, further demonstrating that it is non-Hermiticity, not parity-breaking, that drives the system away from equilibrium.
}

Our goal is to compute the average LEPR through Eq.\,\eqref{eqn:LEPR_small_ah_generic}.
First, let us find the covariance matrix, which at leading order in the non-Hermitian part is given by Eq.\,\eqref{eqn:covariance_matrix_small_ah} and can be found by solving the relation
\begin{equation}
    \underline{\mathcal{A}}_{\mathrm{K}}^{\mathrm{H}} \, \underline{\mathcal{C}} = \: { {-}} \mathds{1} \,.
\end{equation}
In other words, the covariance matrix is the Green's function of the operator ${ {-}}\underline{\mathcal{A}}_{\mathrm{K}}^{\mathrm{H}}$.
Though hard to compute, this Green's function can be found analytically. The explicit solution was obtained in Ref.\,\cite{Ai:2024taz} and reads
\begin{equation}
    \label{eqn:Greens_fnct_ansatz}
    \mathcal{C}(\vec x, \vec x') = \: \int \frac{d^{d-1} \vec{q}_\parallel}{(2\pi)^{d-1}} e^{-i\vec q_\parallel \cdot (\vec x_\parallel - \vec x'_\parallel)} F_{q_\parallel} (w,w') \,,
\end{equation}
where $(w,w')=(\tanh\alpha x_1, \tanh\alpha x'_1)$, $\vec q_\parallel$ is the momentum parallel to the domain wall and orthogonal to $x_1$, and 
\begin{align} 
     F_{q_\parallel} ( w,w')  =  \frac{1}{\alpha\mu} \frac{1}{2\nu} \Bigg[& \theta(w-w') \left(\frac{1-w}{1+w}\right)^{\frac{\nu}{2}} \left(\frac{1+w'}{1-w'}\right)^{\frac{\nu}{2}} \times \notag \\ 
    &\, \times \left(1-3\frac{(1-w)(1+\nu+w)}{(1+\nu)(2+\nu)}\right) \left( 1 - 3\frac{(1-w')(1-\nu + w')}{(1-\nu)(2-\nu)}\right) \notag\\
    & \qquad  + ( w \leftrightarrow w') \Bigg]\,,
    \label{eqn:non-subtracted_kink_green_fnct}
\end{align}
having defined~$\nu = \sqrt{4+ q_\parallel^2/\alpha^2}$.
Note that this Green's function has a singularity for $\nu=2$, i.e. $q_\parallel^2=0$. This points to the presence of the translational zero mode, which arises because translational symmetry is spontaneously broken by the domain wall.
Equation~\eqref{eqn:LEPR_small_ah_generic} on a kink background reads
\begin{align}
    \sigma_\psi(\mathbf{x}) = \:  \mathrm{diag}_{\vec x} \left[ \underline{\mathcal{A}}_{\mathrm{K}}^{\mathrm{AH}} \underline{\mathcal{C}} \left(\underline{\mathcal{A}}_{\mathrm{K}}^{\mathrm{AH}}\right)^\dagger \right] 
    = \:&  4\lambda^2m^2 w^2 \left[ \partial_{\vec{v}}^{(x)} \partial_{\vec{v}}^{(x')} \mathcal{C}(\vec{x},\vec{x}') \right]_{\vec x' = \vec x} \notag \\
    = \: & 4\lambda^2m^2 w^2 \int \frac{d^{d-1}\vec{q}_\parallel}{(2\pi)^{d-1}} \, q_{\mathrm{v}}^2 \, F_{q_\parallel}(w,w)\,,
    \label{eqn:LEPR_DW_with_hom}
\end{align}
where the function $F_{q_\parallel}$ in the coincident limit takes on a compact form
\begin{equation}
    F_{q_\parallel}(w,w) = \: \frac{1}{\alpha\mu}\frac{1}{2\nu} \Bigg[ 1+ 3(1-w^2) \left( \frac{w^2}{\nu^2-1} + \frac{1-w^2}{\nu^2-4} \right) \Bigg] \,.
\end{equation}
Because the system is rotationally symmetric in the $d-1$ directions orthogonal to $x_1$, we can replace $q_{\mathrm{v}}^2$ by $\vec q_\parallel^2/(d-1)$ inside the integral.

Note that the average LEPR in Eq.\,\eqref{eqn:LEPR_DW_with_hom} receives contributions from the large regions where the field is constant and sits in the ordered phase.
To isolate the contribution solely due to the phase boundary, it is convenient to subtract the average LEPR of the uniform phase.
It is irrelevant whether we choose to compute the latter for $\psi_0=\pm m$, since the entropy production is symmetric under $\psi_0\to-\psi_0$.
Although we already have an expression for the average LEPR in the uniform phase, given in Eq.\,\eqref{eqn:average_LEPR_hom}, we need one expressed in the same basis as the one for the domain wall.
Following the same steps as above, we find a similar result up to a different Green's function, which in the coincident limit is given by
\begin{equation}
    F_{q_\parallel}^{\mathrm{hom}} (w,w) = \: \frac{1}{\alpha\mu}\frac{1}{2\nu} \,.
\end{equation}
Then, we can compute the average LEPR solely due to the presence of the phase boundary. We introduce the dimensionless variable $p^2=q_\parallel^2/\alpha^2$ and write
\begin{align}
    \sigma_\psi (\mathbf{x}) = \: & \int_0^\infty d p \ \sigma_p(\tanh \alpha x_1) \,,
    \label{eqn:LEPR_DW}
\end{align}
where
\begin{align}
    \frac{\sigma_p(w)}{\sigma^*} = \: & \frac{4}{(4\pi)^{\frac{d-1}{2}} \Gamma\left(\frac{d-1}{2}\right)} \frac{p^d}{d-1} \frac{3w^2(1-w^2)}{\sqrt{4+p^2}} 
    \left( \frac{w^2}{3+p^2} + \frac{1-w^2}{p^2} \right) \,,
    \label{eqn:LEPR_DW_p_component}
\end{align}
and
\begin{equation}
    \sigma^* = \: \lambda^2 m^2 \alpha^d/\mu \,.
\end{equation}
First, we observe that for any $d\geq2$ the singularity of the function $F_{q_\parallel}$ at $q_\parallel=0$ is regulated by the integral measure, together with the factor of $q_\parallel^2$.
The IR divergence due to the presence of the zero mode disappears in the continuum because the spectral density vanishes fast enough for small momentum.
In a more realistic scenario, where we account for the finiteness of the spatial volume, the spectrum discretises, and the integral is replaced by a sum. Then, one needs to treat the presence of the zero mode with greater care, see for example Refs.\,\cite{Ai:2024taz,Carosi:2024lop,Carosi:2025jpe}.
Next, for large momentum $\sigma_p\propto p^{d-3}$ and, for any $d\geq2$, the LEPR  {$\sigma_\psi$} in Eq.~(\ref{eqn:LEPR_DW}) is once again UV divergent and requires renormalisation.

To get a sense of the LEPR due to the presence of the  {domain wall (DW)}, we plot $\sigma_p$ as defined in Eq.\,\eqref{eqn:LEPR_DW_p_component} for some values of $p=q_\parallel/\alpha$. We set $d=2$ and divide out all parameters since we are mostly interested  in its qualitative behaviour. The result is shown in Figure\,\ref{fig:LEPR_DW_fixed_theta}.
We find that the LEPR is strongly peaked around the DW and is quickly suppressed away from it.
It vanishes at the centre of the DW, where the symmetry is restored, and the field sits in the disordered phase. This is compatible with the LEPR for the uniform phase in Eq.\,\eqref{eqn:average_LEPR_hom}.
Such behaviour aligns with what was found numerically for active phase separation, where it was observed that the LEPR localises at interfaces \cite{CN17, ro2022model, paoluzzi2024noise, Pruessner25}. In those works, the LEPR peaks at the centre of the interface, where the field gradients are largest. In contrast, in the present case, the LEPR vanishes with the order parameter itself and therefore must be zero at the centre of the interface {. Away from the interface we subtracted the uniform background contribution to the LEPR}, giving rise to the double peak shape shown in Figure\,\ref{fig:LEPR_DW_fixed_theta}.

\begin{figure}[t!]
    \centering
    \includegraphics[width=0.55\linewidth]{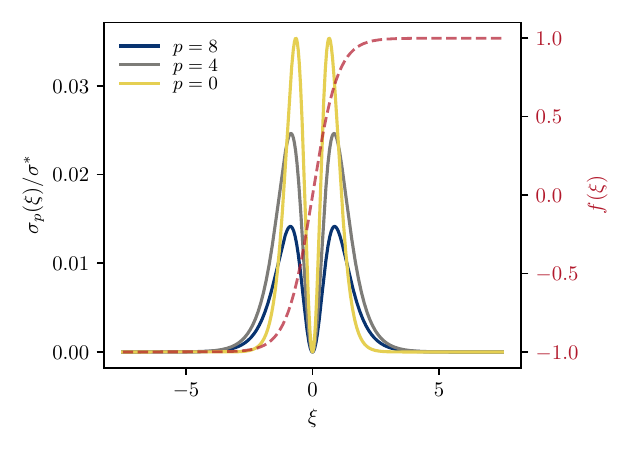}
    \caption{The average LEPR due to the domain wall for $d=2$. The momentum components $\sigma_p$ as defined in Eq.\,\eqref{eqn:LEPR_DW_p_component} are shown for some values of the dimensionless momentum $p=q_\parallel/\alpha$.}
    \label{fig:LEPR_DW_fixed_theta}
\end{figure}

So far, we have studied the particular case in which the vision cone vector is exactly parallel to the domain wall, namely $\mathrm{v}_1=0$.
In the following, we depart from this specific case and study what happens with a generic vision cone vector.
We work in $d=2$, for which all qualitative features of the model are already present.
The coordinates are $(x,y)$, where the domain wall depends on $x$.
We take a generic vision cone vector that lies at some angle $\theta_{\vec v}$ with respect to the $x$ axis, namely
\begin{equation}
    \vec v = \: (\cos\theta_{\vec v},\sin\theta_{\vec v}) \,,
\end{equation}
as shown in the left panel of Figure~\ref{fig:LEPR_DW}. Then, the equation for the domain wall is given by Eq.\,\eqref{eqn:rescaled_deterministic_eom} with $\mathrm{v}_1=\cos\theta_{\vec v}$.
At leading order in the non-reciprocity $\lambda$, or equivalently $\delta$, the EoM remains unchanged and so does the equation for the covariance matrix $\underline{\mathcal{C}}$, so that the solutions of Eqs.\,\eqref{eqn:kink} and~\eqref{eqn:non-subtracted_kink_green_fnct} are still valid.
On the other hand, the anti-Hermitian part of the fluctuation operator now reads
\begin{equation}
    \mathcal{A}_{\mathrm{K}}^{\mathrm{AH}} (\mathbf x, \mathbf x') = \: \delta^{(d)}(\mathbf x-\mathbf x') \; 2\lambda m \tanh \alpha x_1 \left( \cos\theta_{\vec v} \partial_x + \sin\theta_{\vec v} \partial_y\right) \,.
\end{equation}
The average LEPR due to the domain wall at leading order  in $\lambda$ has momentum components given by the following expression
\begin{align}
    \sigma_p(w) = \:  \frac{2}{\pi}\lambda^2m^2 w^2 \alpha^3 \Big\{& p^2 \sin^2\theta_{\vec v}\, F_p(w,w) \notag \\
    &\, - \cos^2\theta_{\vec v} (1-w^2)^2 \left[ \partial_w\partial_{w'} F_p(w,w') \right]_{w'=w} \Big\} \,,
\end{align}
where once again we must subtract the contribution from the uniform regions.
The result is shown in the right panel of Figure~\ref{fig:LEPR_DW} for a reference dimensionless momentum $p$ and for some values of $\theta_{\vec v}$.
We discover that, as $\theta_{\vec v}$ varies from $\pi/2$ to $0$ and the vision cone vector becomes orthogonal to the domain wall, the entropy production localised at the phase boundary increases.
While this is a leading order   result, corrections to it will be perturbative in $\lambda$ and should not affect the qualitative behaviour.

\begin{figure}[h!]
    \centering
    \includegraphics[width=1\linewidth]{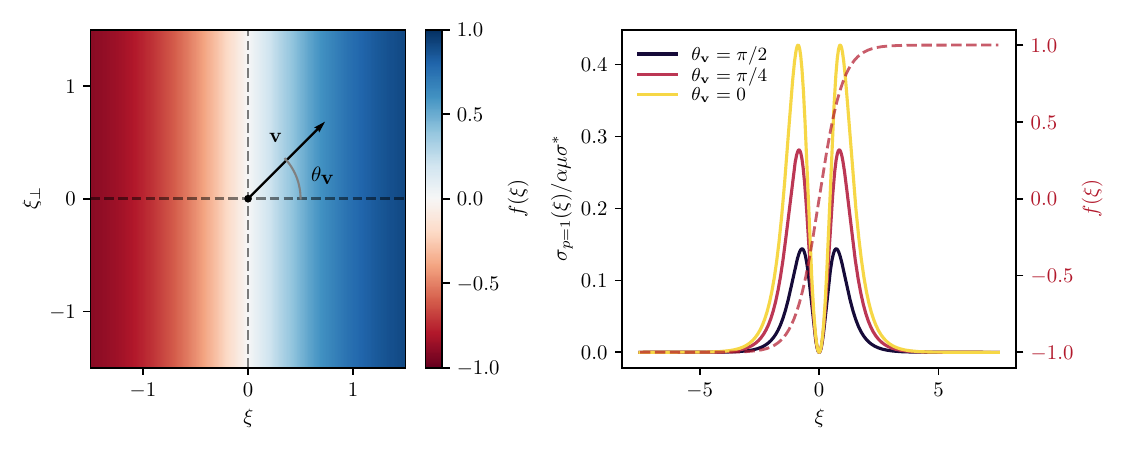}
    \caption{
    \textbf{Left:} Schematic representation of the domain-wall and vision-cone set-up. 
    \textbf{Right:} The average LEPR due to the presence of the domain wall for various values of the tilt angle $\theta_{\vec v}$ {, for fixed $p=1$.}} 
    \label{fig:LEPR_DW}
\end{figure}

\section{Conclusions}
\label{sec:conclusion}

In this work we developed a general framework to quantify time-reversal symmetry breaking in scalar Model A field theories 
 {that exhibit a non-Hermitian linearised dynamics once expanded around a deterministic background.}
 {We reviewed the calculation of the violations of the fluctuation-dissipation theorem within the response-field formalism and using a small-noise expansion around a uniform solution of the deterministic equation of motion.
We further established a local Harada-Sasa relation, which expresses the local entropy production rate (LEPR) in terms of the violations of the fluctuation--dissipation theorem.
This local relation is valid for any background field configuration and does not rely on the small noise expansion.
}
Within this approach, we showed that the departure from equilibrium is fully characterised by the anti-Hermitian component of the   {fluctuation} operator. That said, the addition of any non-conservative term in the field dynamics, such as KPZ-like terms as the ones included in some active field theories of phase separation \cite{cates2025active}, does not break time-reversal symmetry at the level of a small noise expansion around a uniform background. Only anti-Hermitian components do, at first order in such an expansion. 

Our analysis establishes two main results. First, around  {uniform} stationary solutions, the non-Hermitian part of the dynamics generates linear corrections to the FDT, providing a direct and sensitive diagnostic of irreversibility at the level of correlations and response functions. Second, the EPR arises at quadratic order in the non-Hermiticity, reflecting its role as a scalar measure of time-reversal symmetry breaking. This distinction highlights that FDT violations and EPR probe complementary aspects of non-equilibrium dynamics, connected through Harada--Sasa type of relations. We further derived a general expression for the local EPR directly from the Langevin formulation, which is valid for arbitrary  {stationary} reference states within the small-noise regime,  {provided that the non-Hermitian component of the fluctuation operator is small, and its Hermitian component commutes with the noise correlation operator}.

To illustrate the applicability of the framework, we considered a minimal non-Hermitian extension of the Ginzburg–Landau  theory motivated by coarse-grained descriptions of non-reciprocal spin systems. For this model, we obtained exact expressions of the local EPR in both uniform and non-uniform backgrounds. In particular, we found that entropy production vanishes in symmetric phases and becomes localised at interfaces. This result provides an explicit analytical example supporting previous numerical observations in active field theories, where irreversibility concentrates at boundaries, such as domain walls or interfaces.

More broadly, our results extend previous approaches developed for Hermitian dynamics and provide a systematic framework to analyse irreversibility in non-Hermitian  {scalar} field theories arising from non-reciprocal interactions. Such terms naturally appear in coarse-grained descriptions of active matter, driven  systems and other non-equilibrium many-body systems.
It would be natural to extend the analysis to  conserved dynamics (Model B–type theories), where non-Hermitian operators may lead to qualitatively different spatial distributions of entropy production. Applying this framework to other classes of non-equilibrium field theories, including vectorial order parameters or systems with hydrodynamic couplings, may help clarify  general mechanisms through which non-reciprocity generates macroscopic irreversibility in a broader, somewhat more realistic, context.

\section*{Acknowledgments}
MC gratefully acknowledges DL, his research group, and the University of Barcelona for their warm hospitality and stimulating research environment. MC also acknowledges support from the Internationalisation Funds of the TUM Graduate School, which made the initiation of this collaboration possible. DL thanks Matteo Paoluzzi and Sarah Loos for insightful discussions. DL acknowledges support from the COMPLEXSYS26 program at the Kavli Institute for Theoretical Physics (KITP), supported by NSF grant PHY-2309135. 
MC's work is partially funded by the Fundamental Research Funds for the Central Universities in China (grant No. E5ER6601A2). DL, AG and OG acknowledge MCIU/AEI for financial support under grant agreement PID2022-140407NB-C22 and PID2025-170781NB-C22.  AG acknowledges AGAUR and Generalitat de Catalunya for financial support under the call FI SDUR 2023 Ref.~CCI 2021ES05FPR011. OG acknowledges AGAUR and Generalitat de Catalunya for financial support under the call FI SDUR 2024 Ref.~REU/2207/2024, and MCIU under the call FPU24 Ref.~FPU24/03427.

\pagebreak
\section*{References}

\bibliographystyle{iopart-num}

\bibliography{bibliography}

\providecommand{\newblock}{}
\begin{thebibliography}{10}
\expandafter\ifx\csname url\endcsname\relax
  \def\url#1{{\tt #1}}\fi
\expandafter\ifx\csname urlprefix\endcsname\relax\def\urlprefix{URL }\fi
\providecommand{\eprint}[2][]{\url{#2}}

\bibitem{lebowitz1993boltzmann}
Lebowitz J~L 1993 {\em Physics today\/} {\bf 46} 32--38

\bibitem{byrne2022time}
O’Byrne J, Kafri Y, Tailleur J and van Wijland F 2022 {\em Nat Rev Phys\/}
  {\bf 4} 167--183

\bibitem{Elsen18}
Tjhung E, Nardini C and Cates M~E 2018 {\em Phys. Rev. X\/} {\bf 8}(3) 031080
  \urlprefix\url{https://link.aps.org/doi/10.1103/PhysRevX.8.031080}

\bibitem{Li20}
Li Y~I and Cates M~E 2020 {\em J. Stat. Mech.\/} {\bf 2020} 053206
  \urlprefix\url{https://doi.org/10.1088/1742-5468/ab7e2d}

\bibitem{Tailleur08}
Tailleur J and Cates M~E 2008 {\em Phys. Rev. Lett.\/} {\bf 100}(21) 218103
  \urlprefix\url{https://link.aps.org/doi/10.1103/PhysRevLett.100.218103}

\bibitem{Cates15}
Cates M~E and Tailleur J 2015 {\em Annual Review of Condensed Matter Physics\/}
  {\bf 6} 219--244 ISSN 1947-5462
  \urlprefix\url{https://www.annualreviews.org/content/journals/10.1146/annurev-conmatphys-031214-014710}

\bibitem{Solon18}
Solon A~P, Stenhammar J, Cates M~E, Kafri Y and Tailleur J 2018 {\em Phys. Rev.
  E\/} {\bf 97}(2) 020602
  \urlprefix\url{https://link.aps.org/doi/10.1103/PhysRevE.97.020602}

\bibitem{digregorio2018full}
Digregorio P, Levis D, Suma A, Cugliandolo L~F, Gonnella G and Pagonabarraga I
  2018 {\em Phys. Rev. Lett.\/} {\bf 121} 098003

\bibitem{caporusso2020motility}
Caporusso C~B, Digregorio P, Levis D, Cugliandolo L~F and Gonnella G 2020 {\em
  Phys. Rev. Lett.\/} {\bf 125} 178004

\bibitem{shi2020self}
Shi X~q, Fausti G, Chat{\'e} H, Nardini C and Solon A 2020 {\em Phys. Rev.
  Lett.\/} {\bf 125} 168001

\bibitem{caporusso2023dynamics}
Caporusso C~B, Cugliandolo L~F, Digregorio P, Gonnella G, Levis D and Suma A
  2023 {\em Phys. Rev. Lett.\/} {\bf 131} 068201

\bibitem{cates2025active}
Cates M~E and Nardini C 2025 {\em Rep. Prog. Phys.\/} {\bf 88} 056601

\bibitem{Lebowitz99}
Lebowitz J~L and Spohn H 1999 {\em Journal of Statistical Physics\/} {\bf 95}
  333--365

\bibitem{maes2003time}
Maes C and Neto{\v{c}}n{\`y} K 2003 {\em Journal of Statistical Physics\/} {\bf
  110} 269--310

\bibitem{gaspard2004time}
Gaspard P 2004 {\em Journal of Statistical Physics\/} {\bf 117} 599--615

\bibitem{Seifert12}
Seifert U 2012 {\em Rep. Prog. Phys.\/} {\bf 75} 126001
  \urlprefix\url{https://doi.org/10.1088/0034-4885/75/12/126001}

\bibitem{fodor2016far}
Fodor {\'E}, Nardini C, Cates M~E, Tailleur J, Visco P and Van~Wijland F 2016
  {\em Phys. Rev. Lett.\/} {\bf 117} 038103

\bibitem{fodor2022irreversibility}
Fodor {\'E}, Jack R~L and Cates M~E 2022 {\em Annual Review of Condensed Matter
  Physics\/} {\bf 13} 215--238

\bibitem{cocconi2020entropy}
Cocconi L, Garcia-Millan R, Zhen Z, Buturca B and Pruessner G 2020 {\em
  Entropy\/} {\bf 22} 1252

\bibitem{petrelli2020effective}
Petrelli I, Cugliandolo L~F, Gonnella G and Suma A 2020 {\em Phys. Rev. E\/}
  {\bf 102} 012609

\bibitem{cengio2021fluctuation}
Cengio S~D, Levis D and Pagonabarraga I 2021 {\em J. Stat. Mech.\/} {\bf 2021}
  043201

\bibitem{ro2022model}
Ro S, Guo B, Shih A, Phan T~V, Austin R~H, Levine D, Chaikin P~M and Martiniani
  S 2022 {\em Phys. Rev. Lett.\/} {\bf 129} 220601

\bibitem{semeraro2024entropy}
Semeraro M, Negro G, Suma A, Corberi F and Gonnella G 2024 {\em Europhysics
  Letters\/} {\bf 148} 37001

\bibitem{hecht2024define}
Hecht L, Caprini L, L{\"o}wen H and Liebchen B 2024 {\em J. Chem. Phys.\/} {\bf
  161} 224904

\bibitem{di2024variance}
Di~Terlizzi I, Gironella M, Herraez-Aguilar D, Betz T, Monroy F, Baiesi M and
  Ritort F 2024 {\em Science\/} {\bf 383} 971--976

\bibitem{Pruessner25}
Pruessner G and Garcia-Millan R 2025 {\em Rep. Prog. Phys.\/} {\bf 88} 097601
  \urlprefix\url{https://doi.org/10.1088/1361-6633/adff30}

\bibitem{cugliandolo2011effective}
Cugliandolo L~F 2011 {\em J. Phys. A: Math. Theor.\/} {\bf 44} 483001

\bibitem{cugliandolo1997energy}
Cugliandolo L~F, Kurchan J and Peliti L 1997 {\em Phys. Rev. E\/} {\bf 55} 3898

\bibitem{berthier2002nonequilibrium}
Berthier L and Barrat J~L 2002 {\em J. Chem. Phys.\/} {\bf 116} 6228--6242

\bibitem{grigera1999observation}
Grigera T~S and Israeloff N 1999 {\em Phys. Rev. Lett.\/} {\bf 83} 5038

\bibitem{Harada05}
Harada T and Sasa S~i 2005 {\em Phys. Rev. Lett.\/} {\bf 95}(13) 130602
  \urlprefix\url{https://link.aps.org/doi/10.1103/PhysRevLett.95.130602}

\bibitem{JO23}
O'Byrne J 2023 {\em Phys. Rev. E\/} {\bf 107}(5) 054105
  \urlprefix\url{https://link.aps.org/doi/10.1103/PhysRevE.107.054105}

\bibitem{JO24}
O’Byrne J and Cates M~E 2024 {\em Journal of Statistical Mechanics: Theory
  and Experiment\/} {\bf 2024} 113207
  \urlprefix\url{https://doi.org/10.1088/1742-5468/ad8f2b}

\bibitem{JO25}
O’Byrne J and Cates M~E 2025 {\em Journal of Statistical Mechanics: Theory
  and Experiment\/} {\bf 2025} 053204
  \urlprefix\url{https://doi.org/10.1088/1742-5468/add0a2}

\bibitem{hohenberg1977theory}
Hohenberg P~C and Halperin B~I 1977 {\em Rev. Mod. Phys.\/} {\bf 49} 435

\bibitem{tauber2014critical}
T{\"a}uber U~C 2014 {\em Critical dynamics: a field theory approach to
  equilibrium and non-equilibrium scaling behavior\/} (Cambridge University
  Press)

\bibitem{cates2019active}
Cates M~E 2022 {\em Active Field Theories\/} (Oxford University Press) ISBN
  9780192858313
  \urlprefix\url{https://doi.org/10.1093/oso/9780192858313.003.0006}

\bibitem{CN17}
Nardini C, Fodor E, Tjhung E, van Wijland F, Tailleur J and Cates M~E 2017 {\em
  Phys. Rev. X\/} {\bf 7}(2) 021007
  \urlprefix\url{https://link.aps.org/doi/10.1103/PhysRevX.7.021007}

\bibitem{Markovich21}
Markovich T, Fodor E, Tjhung E and Cates M~E 2021 {\em Phys. Rev. X\/} {\bf
  11}(2) 021057
  \urlprefix\url{https://link.aps.org/doi/10.1103/PhysRevX.11.021057}

\bibitem{LC21}
Li Y~I and Cates M~E 2021 {\em J. Stat. Mech.\/} {\bf 2021} 013211
  \urlprefix\url{https://doi.org/10.1088/1742-5468/abd311}

\bibitem{Loos23}
Loos S~A~M, Klapp S~H~L and Martynec T 2023 {\em Phys. Rev. Lett.\/} {\bf
  130}(19) 198301
  \urlprefix\url{https://link.aps.org/doi/10.1103/PhysRevLett.130.198301}

\bibitem{SuchanekPRE2023}
Suchanek T, Kroy K and Loos S~A~M 2023 {\em Phys. Rev. E\/} {\bf 108}(6) 064610
  \urlprefix\url{https://link.aps.org/doi/10.1103/PhysRevE.108.064610}

\bibitem{SuchanekPRL2023}
Suchanek T, Kroy K and Loos S~A~M 2023 {\em Phys. Rev. Lett.\/} {\bf 131}(25)
  258302
  \urlprefix\url{https://link.aps.org/doi/10.1103/PhysRevLett.131.258302}

\bibitem{SuchanekPRE2024}
Suchanek T, Kroy K and Loos S~A~M 2023 {\em Phys. Rev. E\/} {\bf 108}(6) 064123
  \urlprefix\url{https://link.aps.org/doi/10.1103/PhysRevE.108.064123}

\bibitem{paoluzzi2024noise}
Paoluzzi M, Levis D, Crisanti A and Pagonabarraga I 2024 {\em Phys. Rev.
  Lett.\/} {\bf 133} 118301

\bibitem{johnsrud2025}
Johnsrud M~K and Golestanian R 2025 {\em Phys. Rev. Res.\/} {\bf 7}(3) L032053
  \urlprefix\url{https://link.aps.org/doi/10.1103/xx4z-lj5c}

\bibitem{Golestanian25}
Johnsrud M~K and Golestanian R 2025 {\em Phys. Rev. Res.\/} {\bf 7}(3) L032054
  \urlprefix\url{https://link.aps.org/doi/10.1103/flzv-lq7x}

\bibitem{fruchart2026nonreciprocal}
Fruchart M and Vitelli V 2026 {\em arXiv preprint arXiv:2602.11111\/}

\bibitem{you2020nonreciprocity}
You Z, Baskaran A and Marchetti M~C 2020 {\em Proc. Natl. Acad. Sci. U.S.A.\/}
  {\bf 117} 19767--19772

\bibitem{saha2020scalar}
Saha S, Agudo-Canalejo J and Golestanian R 2020 {\em Phys. Rev. X\/} {\bf 10}
  041009

\bibitem{PRXQuantum.4.030328}
Kawabata K, Kulkarni A, Li J, Numasawa T and Ryu S 2023 {\em PRX Quantum\/}
  {\bf 4}(3) 030328
  \urlprefix\url{https://link.aps.org/doi/10.1103/PRXQuantum.4.030328}

\bibitem{gwzr-cjp6}
Hern\'andez-S\'anchez L, Bocanegra-Garay I~A, Ramos-Prieto I, Soto-Eguibar F
  and Moya-Cessa H~M 2025 {\em Phys. Rev. A\/} {\bf 112}(2) L021702
  \urlprefix\url{https://link.aps.org/doi/10.1103/gwzr-cjp6}

\bibitem{PhysRevD.102.125030}
Alexandre J, Ellis J and Millington P 2020 {\em Phys. Rev. D\/} {\bf 102}(12)
  125030 \urlprefix\url{https://link.aps.org/doi/10.1103/PhysRevD.102.125030}

\bibitem{Millington:20228a}
Millington P 2022 {\em PoS\/} {\bf EPS-HEP2021} 735

\bibitem{Garc_s_2025}
Garcés A and Levis D 2025 {\em J. Stat. Mech.\/} {\bf 2025} 043205 ISSN
  1742-5468 \urlprefix\url{http://dx.doi.org/10.1088/1742-5468/adc896}

\bibitem{akritidis2026fate}
Akritidis M, Garc{\'e}s A, Vasilopoulos A, Carosi M, Levis D and Fytas N 2026
  {\em arXiv preprint arXiv:2606.06981\/}

\bibitem{kardar1986dynamic}
Kardar M, Parisi G and Zhang Y~C 1986 {\em Phys. Rev. Lett.\/} {\bf 56} 889

\bibitem{MSR_73}
Martin P~C, Siggia E~D and Rose H~A 1973 {\em Phys. Rev. A\/} {\bf 8}(1)
  423--437 \urlprefix\url{https://link.aps.org/doi/10.1103/PhysRevA.8.423}

\bibitem{de2022path}
De~Pirey T~A, Cugliandolo L~F, Lecomte V and Van~Wijland F 2022 {\em Advances
  in Physics\/} {\bf 71} 1--85

\bibitem{gardiner2010stochastic}
Gardiner C 2010 {\em Stochastic Methods: A Handbook for the Natural and Social
  Sciences\/} Springer Series in Synergetics (Springer Berlin Heidelberg) ISBN
  9783642089626 \urlprefix\url{https://books.google.it/books?id=321EuQAACAAJ}

\bibitem{van2011stochastic}
Van~Kampen N 2011 {\em Stochastic Processes in Physics and Chemistry\/}
  North-Holland Personal Library (North Holland) ISBN 9780080475363
  \urlprefix\url{https://books.google.it/books?id=N6II-6HlPxEC}

\bibitem{onsager1953fluctuations}
Onsager L and Machlup S 1953 {\em Phys. Rev.\/} {\bf 91} 1505

\bibitem{Aron_2010}
Aron C, Biroli G and Cugliandolo L~F 2010 {\em J. Stat. Mech.\/} {\bf 2010}
  P11018 \urlprefix\url{https://dx.doi.org/10.1088/1742-5468/2010/11/P11018}

\bibitem{chow2015path}
Chow C~C and Buice M~A 2015 {\em J. Math. Neurosc.\/} {\bf 5} 8

\bibitem{Thomas_2014}
Thomas P, Fleck C, Grima R and Popović N 2014 {\em J. Phys. A: Math. Theor.\/}
  {\bf 47} 455007
  \urlprefix\url{https://doi.org/10.1088/1751-8113/47/45/455007}

\bibitem{cugliandolo2017rules}
Cugliandolo L~F and Lecomte V 2017 {\em J. Phys. A: Math. Theor.\/} {\bf 50}
  345001

\bibitem{zinn-justin}
Zinn-Justin J 2002 {\em Quantum Field Theory and Critical Phenomena\/} (Oxford
  University Press) ISBN 9780198509233
  \urlprefix\url{https://doi.org/10.1093/acprof:oso/9780198509233.001.0001}

\bibitem{Ai:2024taz}
Ai W~Y, Alexandre J, Carosi M, Garbrecht B and Pla S 2024 {\em JHEP\/} {\bf 05}
  099

\bibitem{Carosi:2024lop}
Carosi M and Garbrecht B 2025 {\em Phys. Rev. D\/} {\bf 111}(8) 085002
  \urlprefix\url{https://link.aps.org/doi/10.1103/PhysRevD.111.085002}

\bibitem{Carosi:2025jpe}
Carosi M 2025 {\em {Quantum fluctuations in cosmological phase transitions:
  from decay rates to the dynamics of expanding bubbles}\/} Ph.D. thesis
  Technical University of Munich

\end{thebibliography}

\end{document}